\documentclass[%
reprint,
amsmath,amssymb,
prstab,
]{revtex4-2}
\usepackage[title]{appendix}
\usepackage{graphicx}
\usepackage{float}
\usepackage{dcolumn}
\usepackage{bm}
\usepackage{gensymb}

\begin{document}

\preprint{APS/123-QED}

\title{An analytical backtracking method for electron beam longitudinal phase space shaping}

\author{N. Sudar and Y. Ding}
\affiliation{SLAC National Accelerator Laboratory, Menlo Park, California 94025, USA
}%

\date{\today}

\begin{abstract}
Many applications of high brightness, highly relativistic electron beams carry strict requirements on longitudinal phase space quality. To meet these requirements, accelerator systems typically utilize dispersive elements to manipulate the correlated energy spread acquired during acceleration or via collective effects. The many free variables in these systems make finding an optimal configuration difficult. We present here an analytical method for tracking a desired final longitudinal phase space backwards through an accelerator section composed of an acceleration or drift section followed by a dispersive section, including analytical models of pertinent collective effects.  This backtracking can serve as a fast and accurate tool for optimization and study of the longitudinal phase space dynamics of an accelerator system.  Here, we consider the LCLS-II accelerator as an example case, using this method to find an accelerator configuration and longitudinal phase space at the injector exit that lead to a significant increase in peak current while maintaining a narrow energy spread at the accelerator exit.  The validity of this method is tested by comparison with Elegant simulations, showing good agreement.
\end{abstract}

\maketitle
\section{Introduction}

The ability to produce high quality, high brightness electron beams has become essential to a wide array of scientific applications.  In the study of ultra-fast science, the X-ray Free Electron Laser (FEL) has become an important tool, requiring a high peak current electron beam with narrow energy spread for succesful operation \cite{RevModPhys.88.015006,FELS2017,EuroXFEL,pile2011first,emma2009first,milne2017swissfel,wang2016soft,hara2013fully,eom2022recent}.  More exotic configurations of the FEL rely on further manipulation of the e-beam current profile \cite{zholents2005method,ATTO,PhysRevLett.125.044801,PhysRevSTAB.15.050707,Kur_2011,zhang2019double,morgan2021attosecond,sudar2020coherent}.  For the study of high energy physics phenomena, linear colliders require high average power beams with low transverse emittance \cite{shiltsev2021modern,michizono2019international,white2018ultra,barklow2020x,nanni2022c}.  Furthermore, many advanced accelerator schemes require high current, high charge drivers and can benefit from specific current profiles \cite{litos2014high,o2016observation,roussel2020single,lemery2015tailored}, potentially serving as building blocks for the next generation of linear colliders and radiation sources \cite{yakimenko2016facet, adli2013beam, emma2022snowmass, emma2021terawatt}.

A common design feature of these accelerator configurations is multi-stage bunch compression, integrating several magnetic bunch compressors between accelerating sections in order to produce the desired current profile and energy spread, while reducing deleterious collective effects \cite{di2014electron}.  However, the final beam quality is still limited by correlated non-linearities in the electron beam longitudinal phase space produced by RF curvature\cite{wiedemann2015particle}, higher order compression, and collective effects such as linac cavity wakefields, resistive wall wakefields \cite{WAKE1,WAKE2}, coherent synchrotron radiation (CSR) \cite{CSR1,CSR3}, and longitudinal space charge (LSC)\cite{LSC}.   These non-linearities can be amplified by compression, potentially creating high current spikes where the head and tail of the beam is over or under compressed respectively \cite{MAXIV, TESSA1}.  These current horns lead to reduction of the longitudinal and transverse phase space quality through projected emittance growth and correlated energy spread from the aforementioned collective effects.

There have been many proposed methods to control these correlations \cite{OPT1,OPT12,OPT2,OPT3,OPT5,OPT6,OPT7}.  To adjust second order non-linear compression the second order energy chirp is typically adjusted with harmonic RF cavities \cite{HARM} . It was shown in \cite{TESSA2,ding2019beam,sudar2020octupole}, that additional higher order non-linear compression can be adjusted by inserting octupoles into the bunch compressors.  However, the above methods can still be limited in their ability to shape the final longitudinal phase space and current profile by constraints such as beam losses, available RF power and beamline apertures, as well as the phase space properties of the electron beam coming out of the injector.  

In the design process of the accelerator configuration, the injector section, where space charge dominates, is typically optimized first for a desired transverse emmitance and current.  This injector output is then used as a starting point for downstream optimization.  Adjustment of the incoming longitudinal phase space can be achieved by phase space linearization \cite{deng2014experimental}, collimation \cite{ding2016beam}, and shaping of the photo-cathode laser envelope \cite{penco2014experimental,zhang2020experimental,lemons2022temporal}.  However, the optimal injector longitudinal phase space is unknown and finding a configuration that leads to the desired final beam properties can be an arduous task due to the many free variables throughout the accelerator and the computational time associated with the fidelity needed to simulate collective effects in particle based codes.  

Here we propose a method to simplify this problem, developing an analytical framework for tracking the longitudinal phase space and current profile at the exit of the accelerator \textit{backwards} through multiple stages of bunch compression, following a similar approach to \cite{TESSA2,floettmann2001generation,zagorodnov2011semianalytical,PhysRevSTAB.9.120701}.  For this backtracking approach, we consider the limit of negligible uncorrelated energy spread, describing the longitudnal phase space and current profile at the accelerator exit as $N^{th}$ order polynomials.  This simplification allows us to track the transformation of the polynomial coefficients through each stage of bunch compression including analytical expressions for pertinent collective effects as the current profile evolves. Our goal here is to use the backtracking approach to develop a fast method to find an accelerator configuration and target longitudinal beam parameters at the injector exit that lead to a desired final longitudinal phase space and current profile, considering terms to arbitrary order.  

In Section II we derive the transformation of the longitudinal phase space and current profile, backtracking through a single bunch compressor.  Discussion on the inclusion of collective effects is given in Section III.  In Section IV we apply the method outlined in Section II and III to the LCLS-II superconducting linac, finding an accelerator configuration and longitudinal distribution at the injector exit that gives a 2 kA peak current beam with $0.1\%$ energy spread at the undulator entrance.  Here we include comparison with forward tracking in Elegant \cite{osti_761286}, demonstrating the viability of the backtracking approach.  A second LCLS-II example case is given in Appendix A, reaching a 4 kA peak current and $0.1\%$ energy spread at the undulator entrance in Elegant simulations, while highlighting some limitations of the proposed method.  Further discussion of the CSR wake derivation is given in Appendix B.

\section{Backtracking through a single bunch compressor}

We start out with a given longitudinal phase space at the exit of the accelerator described by the bunch coordinate, $s_f$ and the energy detuning, $\eta_f = \frac{\gamma-\gamma_0}{\gamma_0}$ where $\gamma_0$ is the energy at $s_f=0$.  We assume a delta function uncorrelated energy distribution with correlated energy chirp given by the Nth order polynomial:
\begin{equation}
\begin{aligned}
&\eta_F(s_f) = h_{f1}s_f+h_{f2}{s_f}^2+h_{f3}{s_f}^3+\ldots+h_{fN}{s_f}^N
\end{aligned}
\end{equation}
We also describe the current profile as an Nth order polynomial distribution with the bunch head at $s_f=S_1$ and tail at $s_f=S_2$:
\begin{equation}
\begin{aligned}
&I_f(s_f) = I_{f0}(1+I_{f1}s_f+I_{f2}{s_f}^2+\\
&I_{f3}{s_f}^3+\ldots+I_{fN}{s_f}^N) \quad S_1 < s_f < S_2 \\
&I_f(s_f) = 0 \quad s_f < S_1\quad \& \quad s_f >S_2
\end{aligned}
\end{equation}

\begin{figure}[h]
\centering
\includegraphics[scale=0.37]{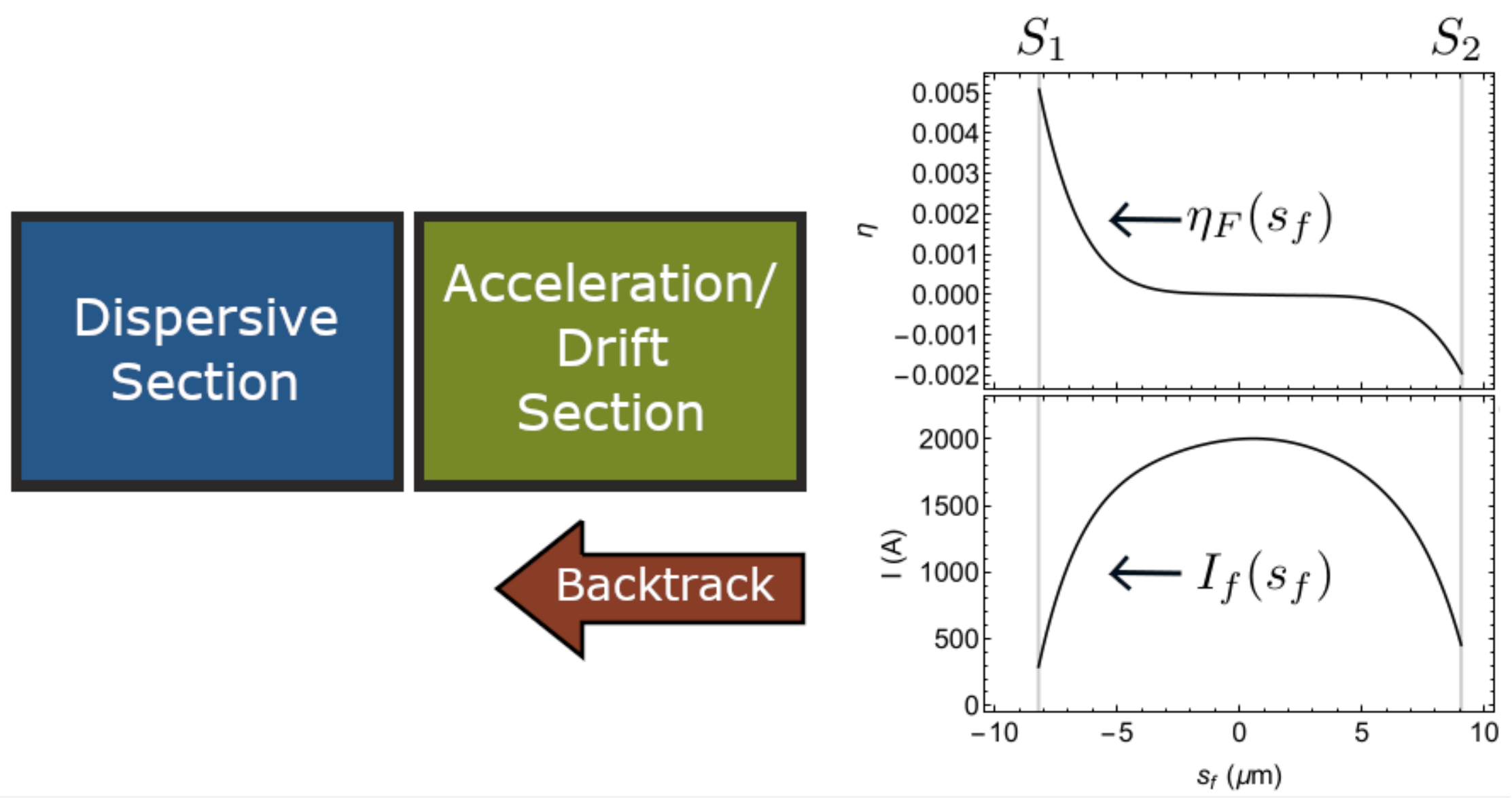}
\caption{Diagram of the backtracking scheme, showing example polynomials, $\eta_F(s_f)$ and $I_f(s_f)$ with endpoints $S_1$ and $S_2$. }
\label{btrack_diagram}
\end{figure}

Here we consider a section of the accelerator composed of a dispersive section (i.e. a bunch compressor) followed by an acceleration section or drift in the forward direction, Fig. \ref{btrack_diagram}.  Our aim is to find how the current profile and chirp at the entrance of the dispersive section depends on the polynomial coefficients of the chirp and current profile at the exit of the acceleration/drift section.

We first track backwards from the exit of the acceleration/drift section to the exit of the dispersive section: 
\begin{equation}
\begin{aligned}
&s_F=s_f\\
&\eta_F=h_{f1}s_f+h_{f2}{s_f}^2+h_{f3}{s_f}^3+\ldots+h_{fN}{s_f}^N\\
&s_2=s_F\\
&\eta_2=\frac{1}{R_{66}}\bigg[\eta_F-H_1 s_F-H_2 {s_F}^2-H_3 {s_F}^3-\ldots-H_N {s_F}^N \bigg]\\
\end{aligned}
\end{equation}
Here, $H_{n}$ describes the chirp acquired by the beam in the acceleration/drift section from RF curvature, collective effects or both.  The change in central energy is given by, $R_{66} = \frac{\gamma_i}{\gamma_f}$. We assume the current profile does not evolve in this section. Now, tracking backwards from the exit to the entrance of the dispersive section gives:
\begin{equation}
\begin{aligned}
&s_I=s_2-D_1 \eta_2 -D_2 {\eta_2}^2-D_3 {\eta_2}^3-\ldots-D_N{\eta_2}^N\\
&\eta_I=\eta_2\\
\end{aligned}
\end{equation}

Here, $D_n$, are the polynomial coefficients describing the longitudinal dispersion in the section. The above transformation gives the chirp at the entrance of the bunch compressor in terms of $s_f$.  In order to write the chirp at the dispersive section entrance, $\eta_I(s_i)$, we must first find $s_f(s_i)$.  For non-trivial polynomial order, the equation $s_I(s_f)=s_i$ can not be inverted in general.  In order to approximate this inversion we can track forwards from the entrance of the dispersive section to the accelerator/drift section exit, writing the chirp at the entrance of the dispersive section in terms of some unknown polynomial coefficients, $h_{In}$:

\begin{equation}
\begin{aligned}
&\bar{s}_I=s_i\\
&\bar{\eta}_I=h_{I1} s_i+h_{I2} {s_i}^2+h_{I3} {s_i}^3+\ldots+h_{IN} {s_i}^N\\
&\bar{s}_2=s_I+D_1 \bar{\eta}_I +D_2 {\bar{\eta}_I}^2+D_3 {\bar{\eta}_I}^3+\ldots+D_N {\bar{\eta}_I}^N\\
&\bar{\eta}_2=\bar{\eta}_I\\
&\bar{s}_F =\bar{s}_2\\
&\bar{\eta}_F= R_{66} \bar{\eta}_2+H_1 \bar{s}_2+H_2 {\bar{s}_2}^2+H_3 {\bar{s}_2}^3+\ldots+H_N {\bar{s}_2}^N\\
\end{aligned}
\end{equation}

Here the $\bar{X}$ notation is simply to differentiate between forward and backtracking.  Inserting $\bar{s}_F(s_i)$ from forward tracking into $\eta_I(s_f)$ from backtracking, we obtain an expression for the chirp at the bunch compressor entrance, $\eta_I(s_i)$, in terms of the unknown initial chirp coefficients, $h_{In}$.  In order to solve for these chirp coefficients in terms of the accelerator parameters and final chirp coefficients, we equate $h_{In}$ to the $n^{th}$ order polynomial coefficient of $\eta_I(s_i)$.

In order to write down the solution to this system of equations, it is useful to define the chirp at the dispersive section exit, $Y_n$, the decompression factors, $d_n$, and the scaled decompression factors, $\bar{d}_n$:
\begin{equation}
\begin{aligned}
&Y_0=0\\
&Y_n=\frac{1}{R_{66}}(h_{fn}-H_n)\\
&d_n=\frac{1}{n!}\frac{\partial^{(n)}}{\partial {s_f}^n}s_i(s_f)\bigg|_{s_f =0}=\delta_{n,1}-D_1 Y_n-\sum_{j=2}^n D_j Q_{n,j}\\
&Q_{n,j}=\\
&\sum_{x_1=0}^{n-1}\sum_{x_2=0}^{x_1-1}\cdots \sum_{x_{j-1}=0}^{x_{j-2}-1}Y_{n-x_1}Y_{x_1-x_2}\cdots Y_{x_{j-2}-x_{j-1}}Y_{x_{j-1}}\\
&\bar{d}_1 = \frac{1}{d_1} \quad\quad \bar{d}_n=-{(\bar{d}_1)}^n d_n\\
\end{aligned}
\end{equation}
Here $\delta_{n,m}$ is the Kronecker delta function. Making use of the above definitions, the initial chirp coefficients can be written in a somewhat compact form:
\begin{equation}
\begin{aligned}
&h_{I(n)}=Y_{n}T_{n,1,0}+\sum_{m=2}^n \sum_{j=1}^{m-1} Y_{n-m+1}T_{n-m+1,m,j}\\
&T_{n,1,0}={(\bar{d}_1)}^n\\
&T_{n,m,j}=\\
&\frac{n}{j!}\frac{(m-2+j+n)!}{(m-1+n)!}{(\bar{d}_1)}^n \sum_{x_1 = 2}^{N_1 - 1}\cdots \sum_{x_{j-1}=2}^{N_{j-1}-1}\bar{d}_{x_1} \cdots \bar{d}_{x_{j-1}} \bar{d}_{N_{j}}\\
&N_l=m+l-1-\sum_{k=1}^{l-1}x_k
\end{aligned}
\end{equation}
Inserting these expressions into Eq. 5 we obtain $\bar{s}_F(s_i)$ in terms of the free variables of the accelerator and the chirp at the accelerator exit.

In the limit of negligble uncorrelated energy spread the transformation of the distribution function of the beam is given in terms of the transformation of the longitudinal phase space coordinates, $I_f[s_F]\delta[\eta_F]\rightarrow I_f[\bar{s}_F(s_i,\eta_i)]\delta[\bar{\eta}_F(s_i,\eta_i)]$, where $\delta[x]$ is the dirac delta function. The current profile at the entrance of the bunch compressor is then given by: 
\begin{equation}
I_i (s_i)=\int_{-\infty}^{\infty} d\eta_i I_f[\bar{s}_F(s_i,\eta_i)]\delta[\bar{\eta}_F(s_i,\eta_i)]
\end{equation}
It was shown in appendix A of \cite{sudar2020octupole} that the above integral can be written as: 
\begin{equation}
I_i (s_i)=\sum_j \frac{I_f[s_f]}{|\frac{d s_I}{ds_f}|}\bigg|_{s_f = s_{f(j)}(s_i)}
\end{equation}

Where $s_{f(j)}(s_i)$ are given by the roots of $\bar{s}_F(s_i)=s_f$. Here we consider solutions that give only one positive root, implying that the longitudinal phase space is piece wise continuous throughout, i.e. no current horns.  In this case, $s_{f(j)}(s_i)=\bar{s}_{F}(s_i)$ and we can drop the sum and absolute value.  The current profile is then defined in the region,  $s_I(S_1)< s_i < s_I(S_2)$ by:

\begin{equation}
\begin{aligned}
&I_i(s_i)=\frac{I_f[\bar{s}_{F}(s_i)]}{\frac{ds_I}{ds_f}[\bar{s}_{F}(s_i)]}\equiv \frac{\sum_{n=0}^{N} \bar{I}_{f(n)}{s_i}^n}{\sum_{n=0}^{N-1} \bar{ds}_{(n)}{s_i}^n}\\
\end{aligned}
\end{equation}
Where we have defined the numerator and denominator in terms of the polynomial coefficients, $\bar{I}_{f(n)}$ and $\bar{ds}_{(n)}$.  Expanding Eq. 10 gives the transformed current profile as a polynomial distribution.  In order to write out this expansion, we first define the compression factors, $C_n$:
\begin{equation}
\begin{aligned}
&C_n=\frac{1}{n!}\frac{\partial^{(n)}}{\partial {s_i}^n}s_f(s_i)\bigg|_{s_i =0}=\delta_{n,1}+D_1 h_{I(n)}+\sum_{j=2}^n D_j q_{n,j}\\
&q_{n,j}=\\
&D_j\sum_{x_1=0}^{n-1}\sum_{x_2=0}^{x_1-1}\cdots \sum_{x_{j-1}=0}^{x_{j-2}-1}h_{I(n-x_1)}h_{I(x_1-x_2)}\cdots h_{I(x_{j-2}-x_{j-1})}\\
\end{aligned}
\end{equation}
Substituting the expression for $h_{I(n)}$ from Eq. 7 into the above expression, the polynomial coefficients in Eq. 10 can be written in terms of the compression and decompression factors as:
\begin{equation}
\begin{aligned}
&\bar{ds}_{(0)}=d_1 \quad\quad \bar{ds}_{(n)}=\sum_{m=2}^{n+1} md_m P_{n,m-1}\\
&\bar{I}_{f(0)}=I_{f0} \quad\quad \bar{I}_{f(n)}=\sum_{m=1}^{n}I_{f0}I_{f(m)}P_{n,m}\\
&P_{n,1}=C_n\\
&P_{n,m}=\\
&\sum_{x_1=1}^{n-1}\sum_{x_2=1}^{x_1-1}\cdots\sum_{x_{m-1}=1}^{x_{m-2}-1}C_{n-x_1}C_{x_1-x_2}\cdots C_{x_{m-2}-x_{m-1}}C_{x_{m-1}}
\end{aligned}
\end{equation}
Making use of the expressions in Eq. 11 and 12, we now expand the denominator of Eq. 10:
\begin{equation}
\begin{aligned}
&\bigg[\frac{1}{n!}\frac{\partial^{(n)}}{\partial {s_i}^n}\bigg(\frac{1}{\frac{ds_i}{ds_f}[s_{f}(s_i)]}\bigg)\bigg]\bigg|_{s_i = 0}\equiv \frac{1}{d1}\delta_{n,1}+\sum_{m=1}^{n} \frac{{V}_{n,m}}{d_1}\\
&{V}_{n,1}=-\frac{\bar{ds}_n}{d_1} \\
&{V}_{n,m}=(-1)^m \bigg(\frac{1}{d_1}\bigg)^{m}\times\\
&\sum_{x_1=1}^{n-1}\sum_{x_2=1}^{x_1-1}\cdots \sum_{x_{m-1}=1}^{x_{m-2}-1}\bar{ds}_{n-x_1} \bar{ds}_{x_1-x_2}\cdots \bar{ds}_{x_{m-2}-x_{m-1}}\bar{ds}_{x_{m-1}}
\end{aligned}
\end{equation}
Finally, combining the results from Eq. 11-13, the polynomial coefficients of the current profile can be written in a compact form as: 
\begin{equation}
\begin{aligned}
&I_i(s) \equiv I_{i0}\big(1+I_{i1}s_i+I_{i2}{s_i}^2+\ldots+I_{iN}{s_i}^N\big)\\
&I_{i(0)} = \frac{I_{f0}}{d_1}\\
&I_{i(n)}=\sum_{m=1}^n V_{n,m}+\sum_{m=1}^{n} I_{f(m)} P_{n,m}\\
&+\sum_{k=1}^{n-1} \bigg(\sum_{m=1}^{n-k} I_{f(m)} P_{n-k,m}\bigg)\bigg(\sum_{j=1}^k V_{k,j}\bigg)\\
&I_{i(N)}=\sum_{m=1}^{N} I_{f(m)} P_{N,m}\\
&+\sum_{k=1}^{N-1} \bigg(\sum_{m=1}^{N-k}  I_{f(m)} P_{N-k,m}\bigg)\bigg(\sum_{j=1}^k V_{k,j}\bigg)
\end{aligned}
\end{equation}

The endpoints of the transformed current distribution are given by transforming the endpoints, $S_1$ and $S_2$ according to Eq. 6:
\begin{equation}
\begin{aligned}
&S_{1i}=\sum_{n=1}^N d_n {S_1}^n  \quad\quad S_{2i}=\sum_{n=1}^N d_n {S_2}^n
\end{aligned}
\end{equation}

In the case of multi-stage bunch compression, we can repeat the steps given by equations 1-15 to find the chirp and current profile at the entrance of the upstream bunch compressor, with $\eta_I(s_i)$ taking the place of $\eta_f(s_f)$, and $I_i(s_i)$ taking the place of $I_f(s_f)$, using the new current profile to calculate pertinent collective effects.

The above method can be adapted for forward tracking simply by replacing the decompression factors in Eq. 6 and 7 with the compression factors, defining the scaled compression factors in the same fashion and swapping the compression and decompression factors in Eq. 11-13.
 
\section{Chirps and collective effects}

In the previous section, we considered the transport through an acceleration section and/or drift, describing the chirp acquired by the beam with the polynomial coefficients, $H_n$.  Here we describe the calculation of this chirp in terms of the parameters of the accelerator and the coefficients of the final current profile.
\subsection{Chirp from RF curvature}
For an acceleration section the chirp is given by the expansion of:
\begin{equation}
\begin{aligned}
\Delta \eta_a(s) = \frac{e V N_c}{mc^2\gamma_f}\cos{(k s +\phi)}
\end{aligned}
\end{equation}

Where m is the electron mass, e is the magnitude of the electron charge, c is the speed of light, $N_c$ is the number of cavities in the accelerator section, V is the voltage in each cavity, k is the RF frequency and $\phi$ is the acceleration phase, and $\gamma_f = \gamma_i+\frac{e V N_c}{mc^2\gamma_f}\cos{(\phi)}$ is the central energy of the electron beam at the exit of the accelerator section \cite{wiedemann2015particle}.

The chirp coefficients $H_{a (n)}$ are then given by:

\begin{equation}
\begin{aligned}
&H_{a(n)} = \frac{e V N_c}{mc^2\gamma_f} \frac{k^n}{n!} \cos{(\phi+n\frac{\pi}{2})}
\end{aligned}
\end{equation}

\subsection{Chirp from cavity wakefields}

We next consider the chirp coming from longitudinal wakefields from the cavities or from the resistive walls of the beam pipe.  The chirp here is given by the convolution of the beam current distribution with the wake kernel associated with the cavity or beam pipe:

\begin{equation}
\begin{aligned}
&\Delta \eta_w(s) = -\frac{eL}{mc^3\gamma_f}\int_{-S_1}^s ds' I(s')w(s-s') \\
&= -\frac{eL}{mc^3\gamma_f}\int_{-S_1}^s ds' I_0\bigg(1+I_1 s'+I_2 {s'}^2\\
&+\ldots+I_N {s'}^N\bigg)w(s-s')
\end{aligned}
\end{equation}

Where L is the length of the acceleration section or drift. 

Following \cite{bane2007wakefields,WAKE2}, for an RF cavity we consider an expansion of an approximation of the wake kernel of the form:

\begin{equation}
\begin{aligned}
&w(s-s')=\alpha e^{-\beta \sqrt{s}}\\
&w(s-s') \sim \alpha\sum_{j=0}^M \frac{(-1)^j\beta^j}{j!} (s-s')^{j/2}
\end{aligned}
\end{equation}

Here $\alpha$ and $\beta$ are determined by the geometry of the cavity and $M$ can be chosen such that the expansion converges over the bunch length.  The chirp is then given by: 
\begin{equation}
\begin{aligned}
&\Delta \eta_w(s) = -\frac{eL_c N_c}{mc^3\gamma_f}\alpha\times\\
&\sum_{j=0}^M\frac{(-1)^j\beta^j}{j!}\sum_{k=0}^N \int_{S_1}^s ds' \chi_k {s'}^k (s-s')^{j/2}
\end{aligned}
\end{equation}
Where $L_c$ is the length of each cavity, $N_c$ is the number of cavities in the acceleration section, and $\chi_0 \equiv I_0$ and $\chi_{j>0} \equiv I_0 I_j$.  This expression can be expanded about $s=0$ up to Nth order to get the polynomial chirp coefficients from the accelerator cavity wakefields, $H_{w(n)}$.
\begin{equation}
\begin{aligned}
&H_{w(n)}=-\frac{eL_c N_c \alpha}{m c^3 \gamma_f}\sum_{k=0}^N \sum_{j=0}^M \sum_{l=0}^k\frac{2(-1)^{j+l} k! \beta^j \chi_k}{j!l!(k-l)!(2l+j+2)}\times\\
&\frac{(1+\frac{j}{2}+l)!}{(n+l-k)!(1+\frac{j}{2}+k-n)!}(-S1)^{1+\frac{j}{2}+k-n}
\end{aligned}
\end{equation}

\subsection{chirp from resistive wall wakefields}

For the resistive wall wake we consider the AC wake kernel from a round pipe given by:

\begin{equation}
w(s-s')=\frac{Z_0 c}{\pi r^2} e^{-\frac{k_r}{2 Q_r} (s-s')} \cos{[k_r(s-s')]}
\end{equation}
Here $k_r$ and $Q_r$ are fitting parameters as discussed in \cite{bane2004resistive}, and $r$ is the radius of the beam pipe.  The chirp is given by:

\begin{equation}
\begin{aligned}
&\Delta \eta_{rw}(s) = -\frac{eL}{mc^3\gamma_f}\frac{Z_0 c}{\pi r^2}\times\\
&\sum_{k=0}^N\int_{S_1}^s ds' \chi_k {s'}^k e^{-\frac{k_r}{Q_r} (s-s')} \cos{[k_r(s-s')]}
\end{aligned}
\end{equation}
Again this expression can be expanded about $s=0$ up to Nth order to get the polynomial chirp coefficients from the resistive wall wakefields, $H_{rw(n)}$.
\begin{widetext}
\begin{equation}
\begin{aligned}
&F_j=\sum_{k=0}^j \frac{(-1)^k (j-1)!}{(2k)!(j-2k-1)!(2k+1)}{Q_r}^{2k} \quad j > 0 \quad \quad F_j = (1+{Q_r}^2)^j F_{|j|} \quad j < 0 \quad \quad F_0 = 0\\
&G_j=\sum_{k=0}^j \frac{(-1)^k j!}{(2k)!(j-2k)!}{Q_r}^{2k} \quad j \ge 0 \quad \quad G_j = (1+{Q_r}^2)^j G_{|j|} \quad j < 0\\
&H_{rw(n)}=-\frac{eL}{mc^3\gamma_f}\frac{Z_0 c}{\pi r^2}\frac{1}{n!{k_r}^{N-n+1}{Q_r}^{n-1}(1+{Q_r}^2)^{N-n+1}}\bigg(e^{k_r S_1 /Q_r}\sum_{k=0}^N \sum_{j=0}^k\big[ \sin{(k_r S_1)Q_r(k-n-j+1)F_{n-k+j-1}}+\\
&\cos{(k_r S_1)}G_{n-k+j-1}\big](-1)^{n+k+j+1}\frac{k!}{j!}{k_r}^{N-k+j}{Q_r}^{k-j}(1+{Q_r}^2)^{N-n+1}{S_1}^j \chi_k \\
&+\sum_{j=0}^{N-n}(-1)^j (n+j)! {k_r}^{N-n-j}{Q_r}^{j+n}(1+{Q_r}^2)^{N-n-j}G_{j+1}\chi_{n+j}\bigg)
\end{aligned}
\end{equation}
\end{widetext}

\subsection{Chirp from longitudinal space charge}

The chirp from longitudinal space charge (LSC) can be found considering the LSC impedance in free space \cite{saldin2004longitudinal,huang2004suppression}:

\begin{equation}
\begin{aligned}
&Z_{LSC}(k)=\frac{i k Z_0}{\pi \gamma^2}\frac{1-\zeta K_1(\zeta)}{\zeta^2}\\
&Z_{LSC}(k)\sim \frac{i k Z_0}{4\pi \gamma^2}[1.232-2\log(\zeta)]
\end{aligned}
\end{equation}
Where $Z_0$ is the impedance of free space, $\zeta=\frac{k \sigma}{\gamma}$, $\sigma$ is the transverse radius of the beam, $K_1(x)$ is the modified bessel function of the second kind, and we have expanded the impedance to first order in $\zeta$.  The instantaneous change in the beam energy is then given by:

\begin{equation}
\begin{aligned}
&d\gamma_{LSC}(k)=\frac{4 \pi }{I_A}\tilde{I}(k) \frac{Z_{LSC}}{Z_0}\\
&= \frac{ik\tilde{I}(k)}{I_A \gamma^2}[1.232+2\log(\frac{\gamma}{k \sigma})]dz
\end{aligned}
\end{equation}

Here $dz$ is an infinitesimal distance along the accelerator, $\tilde{I}(k)$ is the Fourier transform of the current profile and $I_A$ is the Alfven current.  In order to avoid needing to take the Fourier transform of the current distribution and the inverse Fourier transform of the above expression, we approximate that the logarithmic dependence on, $k$, can be described by an effective wave number, $k_c$ related to the length of the electron bunch.  Defining, $k_c=\frac{4\pi I_0}{Q c}$ we obtain:

\begin{equation}
\begin{aligned}
&d\gamma_{LSC}(s)=\frac{1}{I_A \gamma^2}[1.232+2\log(\frac{\gamma}{k_c \sigma})]\int_{-\infty}^{\infty} dk e^{iks}ik\tilde{I}(k) \\
&d\gamma_{LSC}(s)=\frac{1}{I_A \gamma^2}[1.232+2\log(\frac{\gamma}{k_c \sigma})]\frac{\partial I(s)}{\partial s}dz\\
\end{aligned}
\end{equation}

The chirp from LSC is given by, $\Delta\eta_{LSC}(s)=\frac{1}{I_A {\gamma_f}}\frac{\partial I(s)}{\partial s}\mu_x$, where $\mu_x$ for a drift, $\mu_D$, and for an acceleration section, $\mu_A$, are given by:

\begin{equation}
\begin{aligned}
&\mu_D=\frac{L}{{\gamma_1}^2}[1.232+2\log(\frac{\gamma_1}{k_c \sigma})]\\
&\mu_A=\int_0^L dz\frac{1}{\gamma^2}[1.232+2\log(\frac{\gamma}{k_c \sigma})]\\
&=2L\bigg(\frac{1+\log{(\gamma_1)}}{(\gamma_2-\gamma_1)\gamma_1}-\frac{1+\log{(\gamma_2)}}{(\gamma_2-\gamma_1)\gamma_2}+\frac{1.232+2\log{(\frac{1}{k_c \sigma})}}{2\gamma_1\gamma_2} \bigg)\\
\end{aligned}
\end{equation}

For $\mu_A$, we assume the beam undergoes a linear acceleration over the distance $L$, accelerating the beam from an initial to final energy described by $\gamma_1$ and $\gamma_2$ respectively.  The polynomial chirp coefficents from LSC are then given by:

\begin{equation}
\begin{aligned}
H_{LSC(n)}=\frac{1}{I_A {\gamma_f}}\mu_x (n+1)\chi_{n+1}
\end{aligned}
\end{equation}

Here $\gamma_f$ describes the beam energy at the end of the accelerator section under consideration.

\subsection{Chirp from coherent synchrotron radiation}

In order to find an analytical expression for the coherent synchrotron radiation (CSR) energy modulation generated by the beam passing through a bend magnet, we consider only the short beam-long magnet case where the electron beam experiences energy modulation from the entrance transient, a steady state regime, and the exit transient. This is true in many cases where the electron beam peak current or bend strength is large enough to generate significant energy modulation.  For instance, in the case of the LCLS-II accelerator significant energy modulation from CSR is generated in dog-leg magnets located in the long bypass line after the beam reaches full compression.  Following \cite{saldin1997coherent,stupakov2002csr}, we can approximate the CSR chirp as the bunch passes through a bend. This derivation is shown in Appendix B.  Expanding the expressions for the chirps to Nth order we can write the CSR chirp coefficients as: 

\begin{equation}
\begin{aligned}
&H_{CSR(n)} = -\frac{e(-1)^n}{m c^3 4 \pi \epsilon_0 \gamma}\times\\
&\bigg(\frac{4}{3}\log{(4)}(-1)^n\chi_n +\sum_{k=n}^{N}\sum_{i=0}^{n-1}\Xi_{n,k,i}(-1)^k\chi_k {S_2}^{k-n}+\\
&\sum_{k=n+1}^{N}\Xi_{n,k,n}(-1)^k\chi_k {S_2}^{k-n}\\
&+\sum_{k=0}^{N}\Lambda_{n,k}^0 \Phi {\rho}^{\frac{1}{3}}(-1)^k\chi_k {S_2}^{k-n-\frac{1}{3}}+\\
&\sum_{m=0}^{n}\sum_{k=m+1}^{N}\sum_{i=0}^{m}\Lambda_{n,m,k,i}^1 \Phi {\rho}^{\frac{1}{3}}(-1)^k\chi_k {S_2}^{k-n-\frac{1}{3}}-\\
&\sum_{k=n+1}^{N}\sum_{i=0}^{m}\Lambda_{n,k,i}^2(-1)^k \chi_k {S_2}^{k-n}-\sum_{i=0}^{n-1}\Lambda_{n,n,i}^2(-1)^n\chi_n+\\
&\sum_{k=0}^{N-n}\sum_{m=0}^{n} (-1)^{k+n}\chi_{k+n}{S_2}^k \bigg[ \Upsilon_{n,m,k}^0 +\sum_{i=0}^{m+k-1}\Upsilon_{n,i,m,k}^1 -\\
&\sum_{i=0}^{m+k-1}\sum_{l=0}^{m+k-i-2}\Upsilon_{n,l,i,m,k}^2-\sum_{i=0}^{m+k-1}\sum_{l=m+k-i+1}^{3(m+k)-1}\Upsilon_{n,l,i,m,k}^2\bigg]\bigg)
\end{aligned}
\end{equation}

Here $\Phi$ is the bend angle of the magnet, $\rho$ is the bend radius and $S_2$ is the bunch coordinate of the electron beam tail.  The steady state condition is met when $\Phi > \bigg[ \frac{24}{\rho}(S_2-S_1)\bigg]^{1/3}$.  The coefficients $\Xi$, $\Lambda^0$, $\Lambda^1$, $\Lambda^2$, $\Upsilon^0$, $\Upsilon^1$, and $\Upsilon^2$ are independent of the magnet and electron beam parameters and are shown in Appendix B.

\begin{figure*}[t]
\includegraphics[scale=0.2]{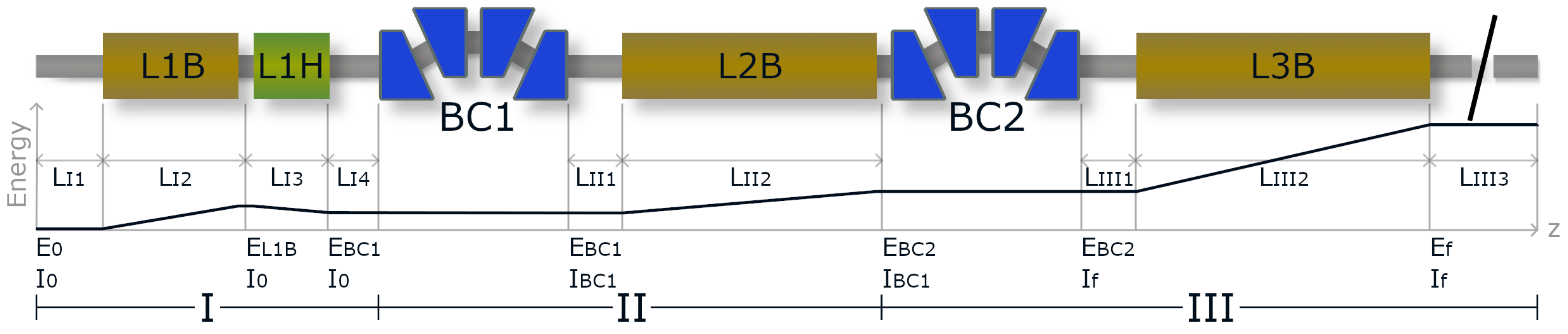}
\caption{Layout of the LCLS-II accelerator, showing the three acceleration sections, L1, L2 and L3, the harmonic linearizer, L1x, the two bunch compressors, BC1 and BC2, and the bypass line following L3.  The energy and current is shown at pertinent locations as well as useful distances for LSC calculations, $L_{In}$, $L_{IIn}$, and $L_{IIIn}$. The bactracking method detailed in section II, can be done iteratively over beamline sections III, II, and I.}
\label{beamlinelayout}
\end{figure*}

\section{LCLS-II example}

Here we consider the LCLS-II accelerator as an example case for making use of the backtracking method, specifically focusing on the hard x-ray line \cite{brachmann2019lcls}.  Downstream of the injector, the LCLS-II beamline consists of an acceleration section, L1B, harmonic cavities, L1H, a 4 dipole chicane bunch compressor, BC1, a second acceleration section, L2B, a second 4 dipole chicane bunch compressor, BC2, a third acceleration section, L3B, and a long bypass line, Fig. \ref{beamlinelayout}. Collective effects in the long bypass line, non-linear compression in the two bunch compressors, and higher order correlations from the injector create difficulties in reaching peak currents of several kA and narrow energy spread, $<0.1 \%$.  Achieving these parameters is important to the efficient operation of the hard X-Ray FEL.  Furthermore, achieving a nearly flattop current distribution is beneficial to the operation of self seeding schemes \cite{ding2016beam}. 

Beginning with a beam satisfying the above conditions, we can backtrack to find the necessary distribution at the exit of the LCLS-II injector.  In order to simplify the problem, we choose $E_0$, $E_{BC1}$, $E_{BC2}$, $E_f$, $I_0$, and $I_{BC1}$ to be near the nominal operating values of the LCLS-II accelerator.  The specified current and chirp polynomial coefficients at the undulator entrance and additional free parameters, $L_{1B}$ voltage, $L_{1B}$ phase, $L_{1H}$ phase, $L_{2B}$ phase, and $L_{3B}$ phase, can be adjusted to achieve a smooth, roughly symmetric current distribution with minimized linear chirp at the injector exit.  For this example case we consider chirp and current terms up to 6th order and dispersive terms up to 3rd order. 

The efficacy of the backtracking method can be checked by comparison with forward tracking in Elegant.  To do this we generate a 6-D phase space from the longitudinal phase space at the injector exit found from backtracking, considering a bi-gaussian transverse distribution with $\epsilon_{nx}=\epsilon_{ny} =$ 0.37 mm-mrad. The bend angles in the bunch compressors, cavity voltages and cavity phases used in the backtracking are used in the Elegant lattice and can be found in Table I. 

\noindent
\begin{table}[ht]
\caption{LCLS-II example case 1 parameters}
\begin{tabular}{|c|c|c|c|}
\hline
\\[-1em]
$L_{I1}$ $(m)$ & 52 & $N_{C}$ $(L1B)$ & 16\\
\hline
\\[-1em]
$L_{I2}$ $(m)$ & 22 & $L_{C}$ $(m)$ & 1.0377\\
\hline
\\[-1em]
$L_{I3}$ $(m)$ & 14 & $\lambda_{C}$ $(m)$ & 0.23061\\
\hline
\\[-1em]
$L_{I4}$ $(m)$ & 7 & $N_{CH}$ $(L1H)$ & 16\\
\hline
\\[-1em]
$L_{II1}$ $(m)$ & 50 & $\lambda_{CH}$ $(m)$ & 0.07687\\
\hline
\\[-1em]
$L_{II2}$ $(m)$ & 155 & $N_{C}$ $(L2B)$ & 96\\
\hline
\\[-1em]
$L_{III1}$ $(m)$ & 34 & $N_{C}$ $(L3B)$ & 160\\
\hline
\\[-1em]
$L_{III2}$ $(m)$ & 253 & $L_{rw}$ (cu)  $(m)$& 339.1\\
\hline
\\[-1em]
$L_{III3}$ $(m)$ & 2920 & $L_{rw}$ (s.s.) $(m)$& 2584.5\\
\hline
\\[-1em]
$E_{0}$ $(MeV)$ & 92 & $I_{0}$ $(A)$ & 11.7\\
\hline
\\[-1em]
$E_{BC1}$ $(MeV)$ & 250 & $R_{56}^{(1)}$ $(mm)$ & -47.37\\
\\[-1em]
\hline
$E_{BC2}$ $(MeV)$ & 1500 & $I_{BC1}$ $(A)$ & 33.3\\
\\[-1em]
\hline
\\[-1em]
$E_{f}$ $(MeV)$ & 4000 & $R_{56}^{(2)}$ $(mm)$ & -43.56\\
\hline
\\[-1em]
$V_{L1B}$ $(MV)$ & 15.992 & $\phi_{L1B}$$\degree$ & -25.06 \\
\hline
\\[-1em]
$V_{L1H}$ $(MV)$ & 4.601 & $\phi_{L1H}$$\degree$ & -175.12 \\
\\[-1em]
\hline
$V_{L2B}$ $(MV)$ & 16.065 & $\phi_{L2B}$$\degree$ & -35.63 \\
\\[-1em]
\hline
\\[-1em]
$V_{L3B}$ $(MV)$ & 15.67 & $\phi_{L3B}$$\degree$ & 0 \\
\hline
\hline
\\[-1em]
$h_{f0}$ & 0 & $I_{f0}$ $(A)$ & 2000 \\
\hline
\\[-1em]
$h_{f1}$ $(m^{-1})$ & -11.91 & $I_{f1}$ & 6428.16 \\
\hline
\\[-1em]
$h_{f2}$ $(m^{-2})$ & 5.39$\times 10^{5}$ & $I_{f2}$ & -5.48$\times 10^9$ \\
\hline
\\[-1em]
$h_{f3}$ $(m^{-3})$ & 3.76$\times 10^{10}$  & $I_{f3}$ & -1.83$\times 10^{14}$ \\
\hline
\\[-1em]
$h_{f4}$ $(m^{-4})$ & 2.86$\times 10^{17}$ & $I_{f4}$ & -5.48$\times 10^{18}$ \\
\hline
\\[-1em]
$h_{f5}$ $(m^{-5})$ & -8.17$\times 10^{22}$  & $I_{f5}$ & 5.58$\times 10^{24}$\\
\hline
\\[-1em]
$h_{f6}$ $(m^{-6})$ & 2.08$\times 10^{27}$ & $I_{f6}$ & -9.88$\times 10^{29}$ \\
\hline
\\[-1em]
$h_{i0}$ & 0 & $I_{i0}$ $(A)$ & 11.7 \\
\hline
\\[-1em]
$h_{i1}$ $(m^{-1})$ & -0.026 & $I_{i1}$ &  -23.43 \\
\hline
\\[-1em]
$h_{i2}$ $(m^{-2})$ & -627.73 & $I_{i2}$ & -53277.4 \\
\hline
\\[-1em]
$h_{i3}$  $(m^{-3})$ & 26168.05 & $I_{i3}$ & 5.49$\times 10^{7}$ \\
\hline
\\[-1em]
$h_{i4}$  $(m^{-4})$ & -1.43$\times 10^{7}$ & $I_{i4}$  & -7.29$\times 10^{10}$ \\
\hline
\\[-1em]
$h_{i5}$  $(m^{-5})$ & 2.65$\times 10^{10}$ & $I_{i5}$ & 1.83$\times 10^{13}$ \\
\hline
\\[-1em]
$h_{i6}$  $(m^{-6})$ & 1.13$\times 10^{12}$ & $I_{i6}$ & -4.68$\times 10^{16}$ \\
\hline
\end{tabular}
\end{table}

\subsection{Region III}

Following Eq. 1-15, we begin backtracking through Region III consisting of the BC2 bunch compressor, L3B linac, and long bypass line as shown in Fig. \ref{beamlinelayout}. The bypass line consists of 3 dog-legs and 6 small chicanes used to compensate for anomalous dispersion.  As stated in section II, we do not consider evolution of the current profile in the bypass line.  This condition limits the correlated energy spread and peak current at the bypass line entrance.  For this example case, we consider a 2 kA peak current.  An additional example where the current profile evolves significantly in the bypass line, reaching a peak current of 4 kA is shown in Appendix A.

The LSC chirp in the bypass line can be calculated from Eq. 29, using $\mu_D$ assuming a beam size throughout of 50 $\mu m$.  The resistive wall wakefield in the bypass line can be calculated from Eq. 24, for a copper pipe with radius 17.4 mm with $k_r=6.0423\times 10^4$ and $Q_r=1.6949$, and a stainless steel pipe with 24.5 mm radius with $k_r=1.3748\times 10^4$ and $Q_r=1.1388$. The dog leg and chicane magnets produce a CSR chirp which can be calculated using Eq. 30.  Magnet strengths are given in Table III in Appendix B. The LSC chirp in L3B can be calculated from Eq. 29, using $\mu_A$ assuming a beam size throughout of 100 $\mu m$. The wakefield in L3B comes from the cavity walls and can be calculated from Eq. 21 using $\alpha = 4.15\times 10^{13}$ and $\beta = 23.973$.  Finally the chirp from RF curvature in L3 can be found from Eq. 17.  

The BC1 and BC2 bunch compressors are both 4 dipole chicanes. The associated dispersive terms are given by:
\begin{equation}
\begin{aligned}
&D_1=R_{56}=-2{\theta_m}^2(L_d+\frac{2}{3}L_m)\\
&D_{(n)} = (-1)^{n+1}\frac{n+1}{2}D_1
\end{aligned}
\end{equation}
Where $L_m$ is the magnet length,  $\theta_m$ is the bend angle, and $L_d$ is the drift length between the first and second, and third and fourth magnets. The required $R_{56}$ of the bunch compressor, BC2, can be found in terms of the total chirp at the BC2 exit, $Y_1$, the peak current at the BC2 exit, $I_{f0}$, and the peak current at the BC2 entrance, $I_{BC1}$, with $R_{56(2)} = \frac{I_{BC1}-I_{f0}}{Y_1 I_{BC1}}$. The non-linear dispersive terms can be found based on Eq. 31.  In order to treat the bunch compressor as a point like transformation, we ignore CSR in the bunch compressor.  A small adjustment to the 2nd order dispersion, $T_{566}\sim -1.41 R_{56}$, can be introduced to compensate for additional non-linear dispersion introduced by CSR energy modulation in the bunch compressor.  With these quantities we can find the current profile and chirp at the entrance of BC2.  Pertinent parameters are shown in Table 1.

\begin{figure}[h]
\centering
\includegraphics[scale=0.45]{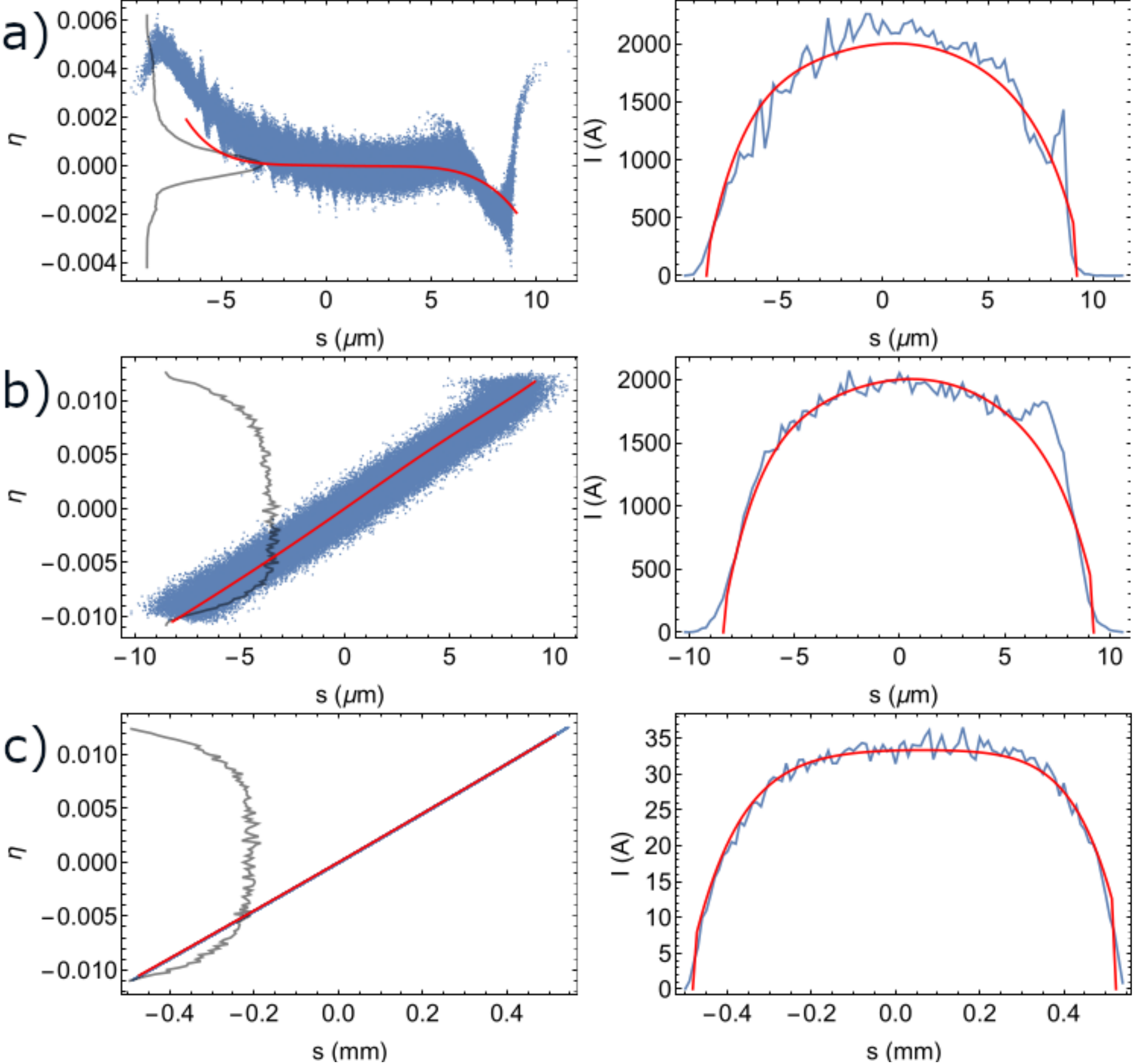}
\caption{Longitudinal phase space (left) and current profile (right) at the undulator entrance (a), BC2 exit (b), and BC2 entrance (c), with polynomials from backtracking (red) and particles from Elegant (blue).}
\label{regionIII}
\end{figure}

\begin{figure}[h]
\centering
\includegraphics[scale=0.42]{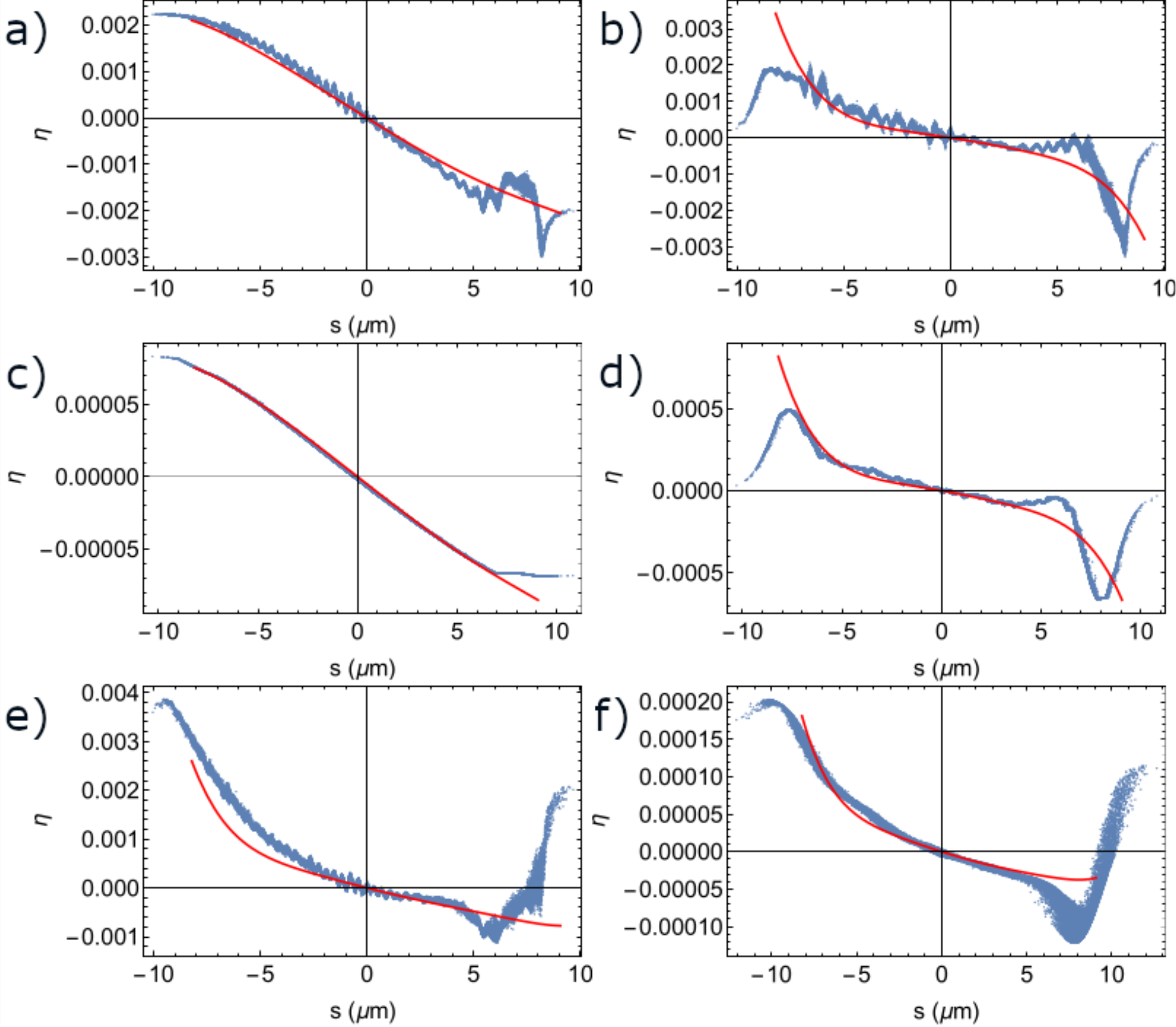}
\caption{Chirp from resistive wall wakefield in bypass line a), LSC in bypass line b), cavity wakefield in L3B c), LSC in L3B d), CSR in the bypass line e), CSR from only the first dog leg magnet f), with polynomials from backtracking (red) and particles from difference between Elegant sims with respective collective effect turned on and off (blue).}
\label{regionIII_ce}
\end{figure}

Figure \ref{regionIII}a shows the longitudinal phase space and current profile at the undulator entrance that is backtracked to the BC2 exit (\ref{regionIII}b), and BC2 entrance (\ref{regionIII}c).   Comparison with the results from forward tracking in Elegant shows good agreement with the major difference found in the chirp at the head of the beam.  This can be attributed to small differences in the CSR and LSC modulations in the bypass line. Figure \ref{regionIII_ce} shows comparison between the analytical estimates of the energy modulation from collective effects and the energy difference in Elegant turning the collective effect on and off in the specific region.

\subsection{Region II}

The current profile and chirp at the entrance of BC2 found from backtracking in Region III can be used to backtrack to the entrance of BC1 again following Eq. 1-15 for Region II of the accelerator.  Calculation of the chirp follows a similar path.  The LSC chirp can be calculated from Eq. 29, using $\mu_D$ over the distance $L_{II1}$ and $\mu_A$ over the distance $L_{II2}$, assuming a beam size throughout of 50 $\mu m$.  The wakefields again come from the cavity walls, identical to those in L3B.  There are no significant sources of CSR.  Again the chirp from RF curvature in L2B can be found from Eq. 17 and the $R_{56}$ of the bunch compressor, BC1, can be found in terms of the total chirp at the BC1 exit, $Y_1$, the peak current at the BC1 exit, $I_{BC1}$, and the peak current at the BC1 entrance, $I_0$, giving $R_{56(1)} = \frac{I_{0}-I_{BC1}}{Y_1 I_{0}}$.  Again the non-linear dispersion terms are found based on Eq. 31.  Ignoring CSR in BC1 does not result in significant differences in the compression.

\begin{figure}[h]
\centering
\includegraphics[scale=0.45]{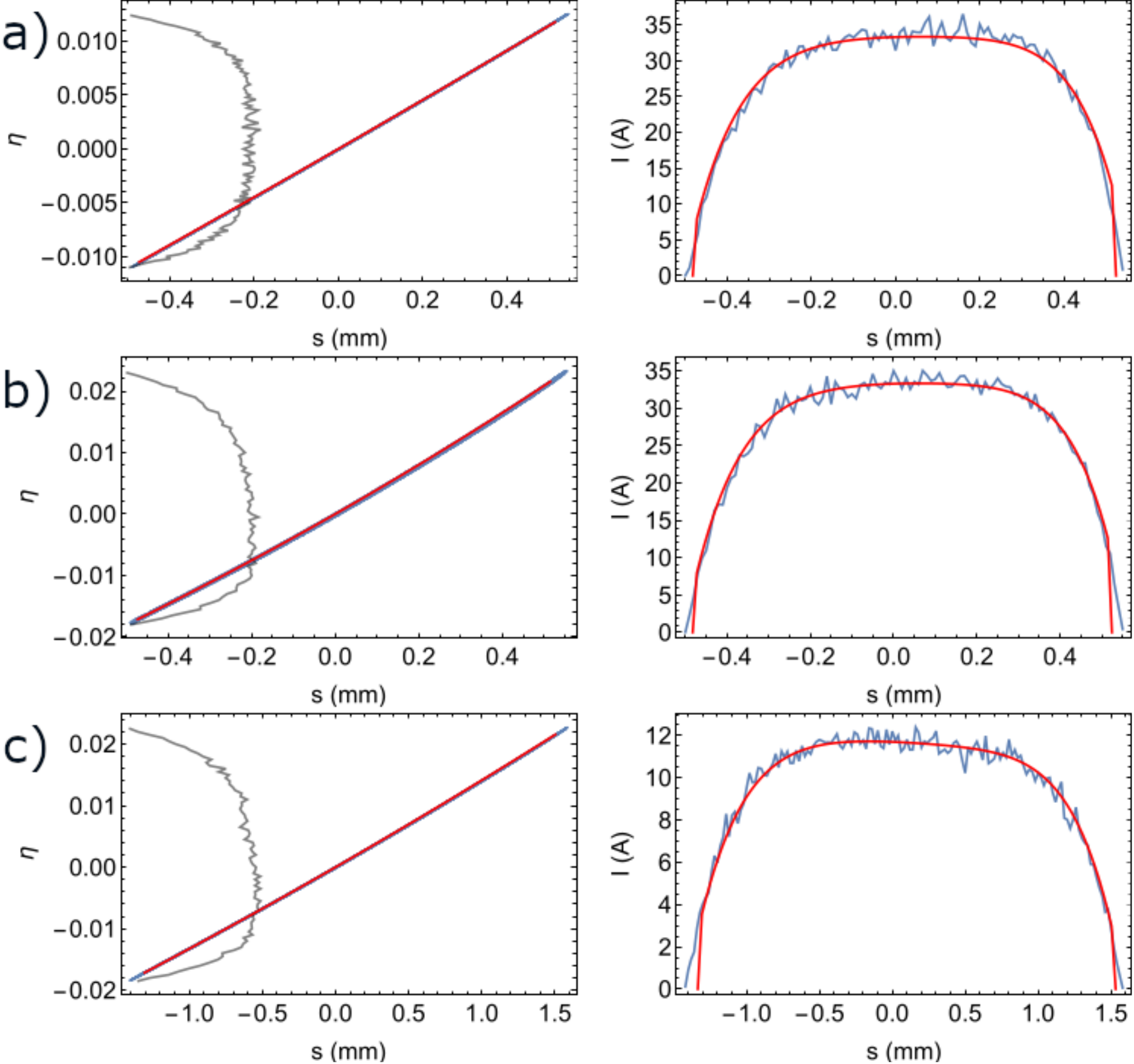}
\caption{Longitudinal phase space (left) and current profile (right) at the BC2 entrance (a), BC1 exit (b), and BC1 entrance (c) with polynomials from backtracking (red) and particles from Elegant (blue).}
\label{regionII}
\end{figure}

\begin{figure}[h]
\centering
\includegraphics[scale=0.42]{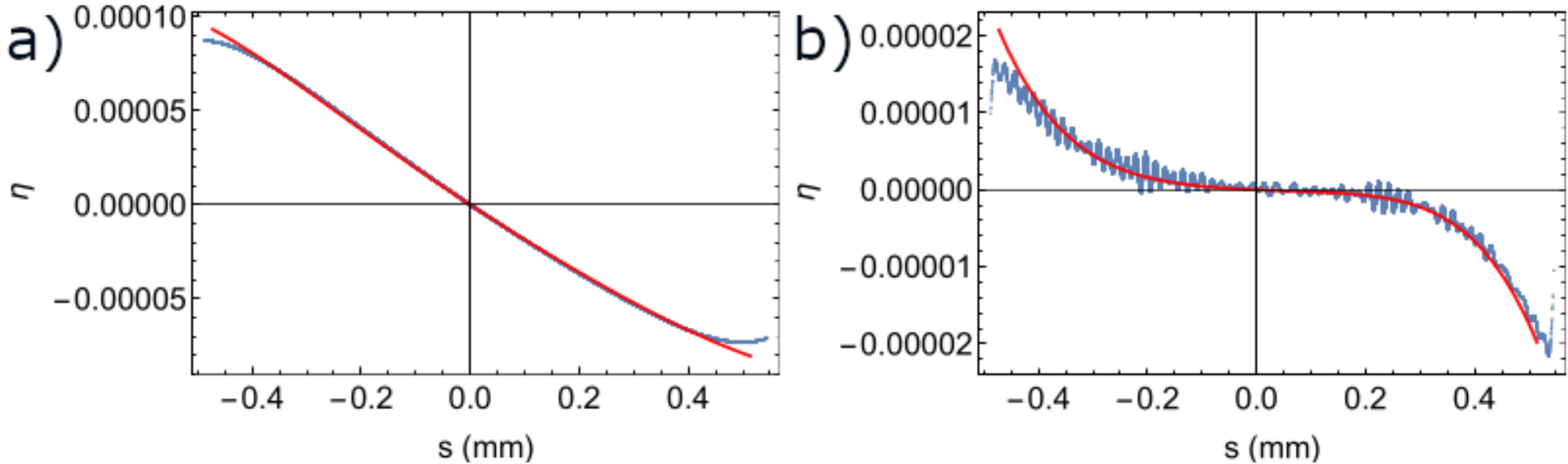}
\caption{Chirp from cavity wakefield in L2B a), LSC in L2B b), with polynomials from backtracking (red) and particles from difference between Elegant sims with respective collective effect turned on and off (blue).}
\label{regionII_ce}
\end{figure}

Figure \ref{regionII}a shows the longitudinal phase space and current profile at the BC2 entrance that is backtracked to the BC1 exit (\ref{regionII}b), and BC1 entrance (\ref{regionII}c).   Comparison with the results from forward tracking in Elegant shows good agreement, with comparison of the analytical estimates and simulated collective effect modulations shown in Fig. \ref{regionII_ce}.

\subsection{Region I}

The current profile and chirp at the entrance of BC1 found from backtracking in Region II can be used to backtrack to the exit of the LCLS-II laser heater, which we consider the exit of the injector.  Since there is no dispersive section, this can be done simply using Eq. 3 to find the change in the chirp in the region. The LSC chirp can be calculated from Eq. 23, using $\mu_D$ over the distance $L_{I1}$ and $L_{I4}$, and $\mu_A$ over the distance $L_{I2}$ and $L_{I3}$, assuming a beam size throughout of 350 $\mu m$.  The wakefields again come from the cavity walls, with the L1B wake identical to L2B and L3B, and L1H wake calculated with $\alpha = 2.3\times 10^{14}$ and $\beta = 34.503$.  There is no significant source of CSR.  Again the chirp from RF curvature in Region I can be found from Eq. 17 for both L1B and L1H. 

\begin{figure}[h]
\centering
\includegraphics[scale=0.45]{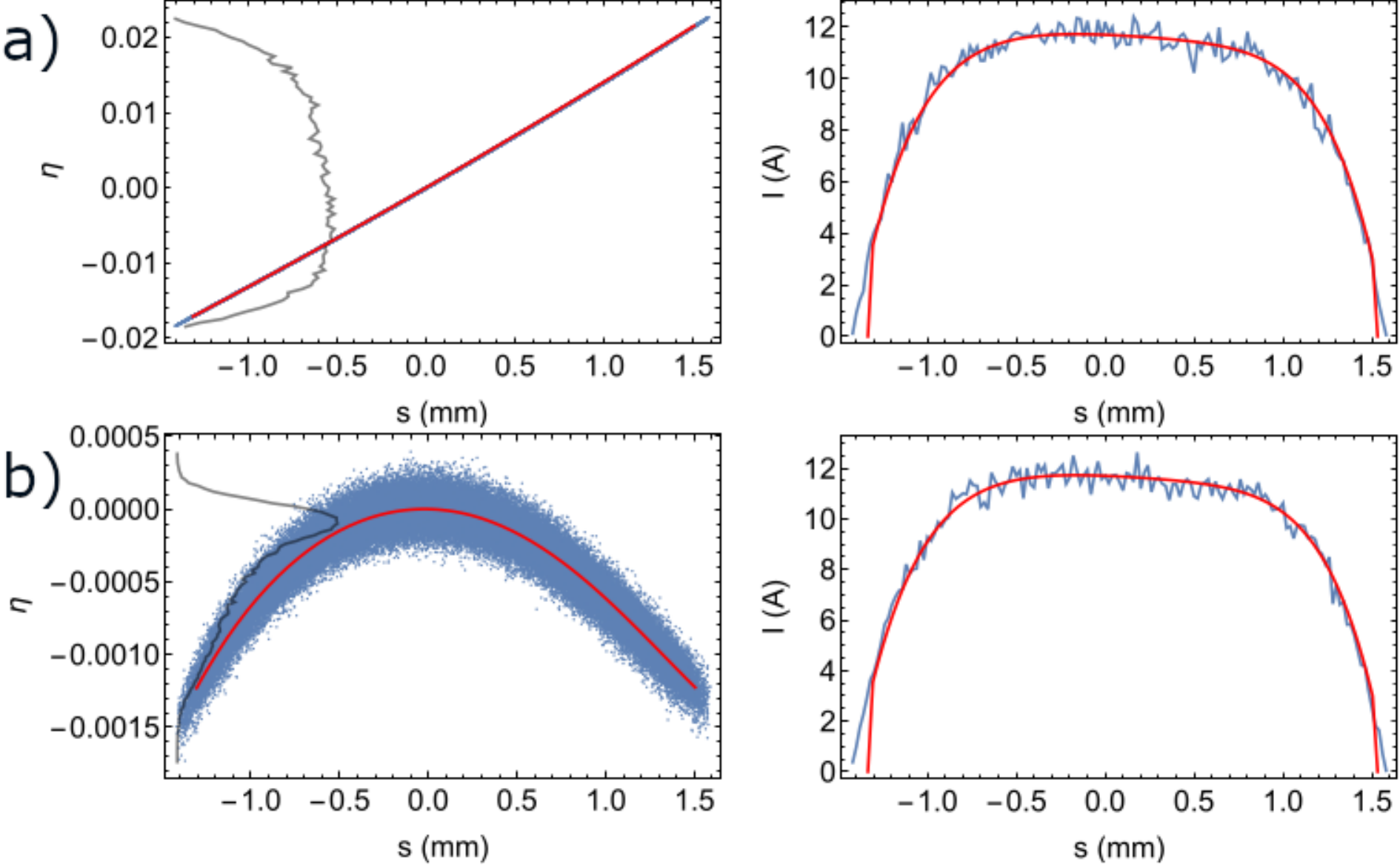}
\caption{Longitudinal phase space (left) and current profile (right) at the BC1 entrance (a), and laser heater exit (b), with polynomials from backtracking (red) and particles from Elegant (blue).}
\label{regionI}
\end{figure}

\begin{figure}[h]
\centering
\includegraphics[scale=0.42]{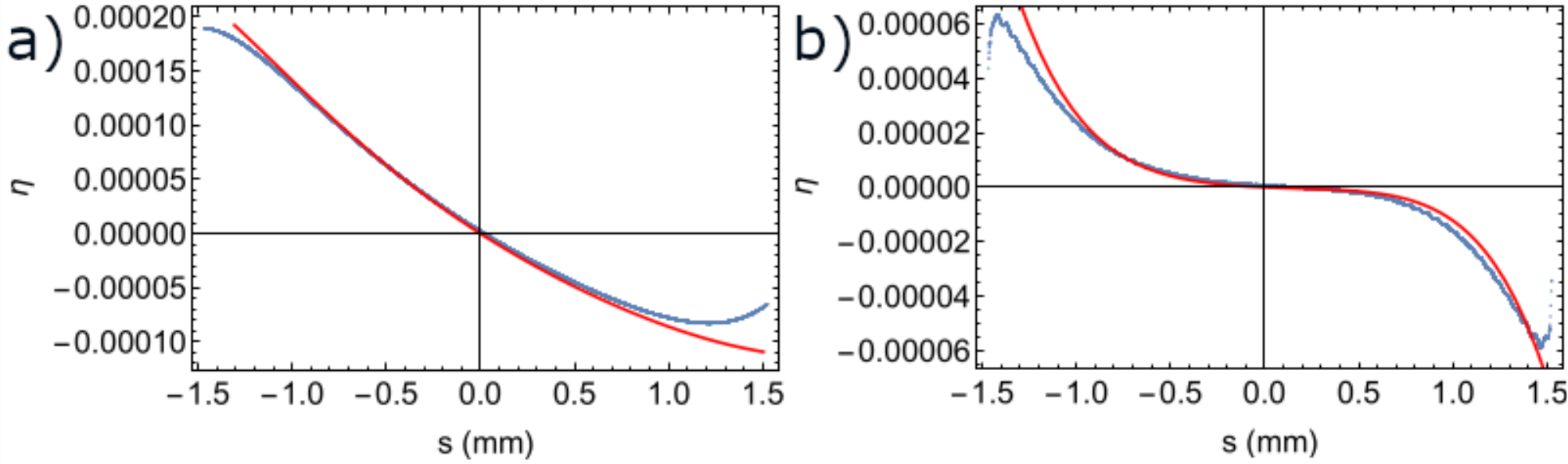}
\caption{Chirp from cavity wakefield in L1B and L1H a), LSC in L1B and L1H b), with polynomials from backtracking (red) and particles from difference between Elegant sims with respective collective effect turned on and off (blue).}
\label{regionI_ce}
\end{figure}

Figure \ref{regionI}a shows the longitudinal phase space and current profile at the BC1 entrance that is backtracked to the laser heater exit (\ref{regionI}b), with the latter used for forward tracking in Elegant.  Here an uncorrelated gaussian energy spread of $\sigma_{\eta} = 0.0001$ is applied to the beam in order to suppress the microbunching instability. Again comparison with the results from forward tracking in Elegant shows good agreement, with comparison of the analytical estimates and simulated collective effect modulations shown in Fig. \ref{regionI_ce}.

\begin{figure}[h]
\centering
\includegraphics[scale=0.46]{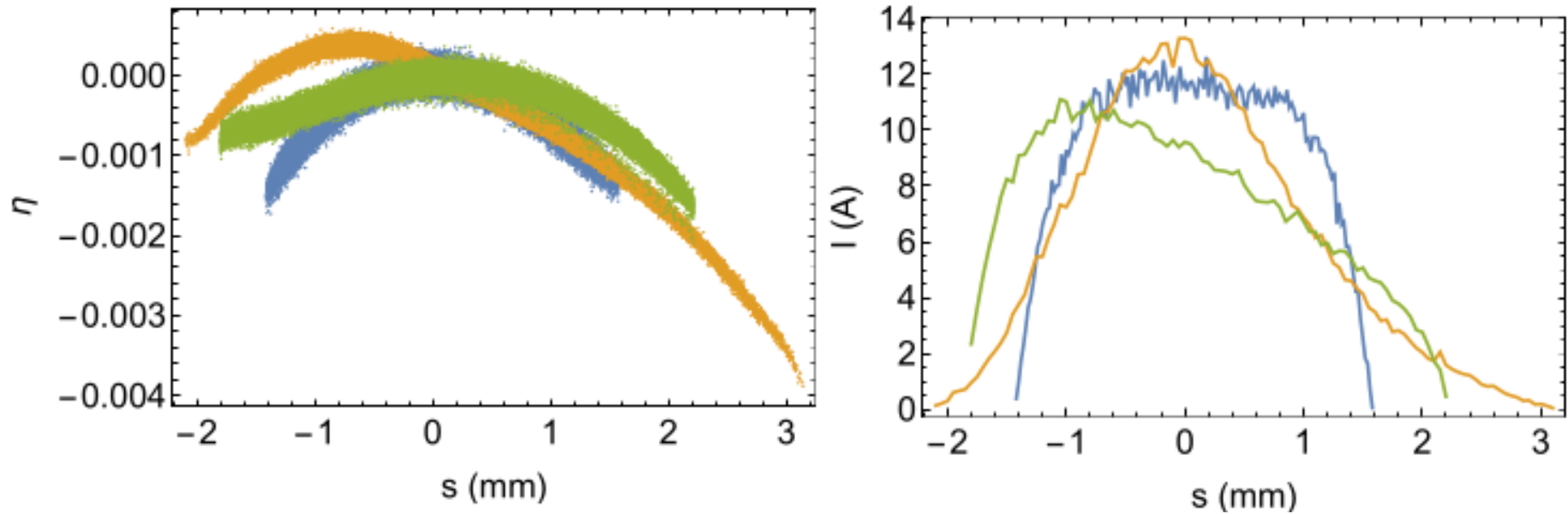}
\caption{Comparison of the longitudinal phase space (left) and current profile (right) at the laser heater exit from backtracking (blue) and nominal LCLS-II cases with flattop (yellow) and gaussian (green) cathode laser temporal profiles.}
\label{PS_comp}
\end{figure}

The ability to produce this longitudinal phase space and current profile at the injector exit requires further study.  Comparing with nominal LCLS-II injector configurations using flattop \cite{marcus2017lcls} and gaussian \cite{neveu2020lcls} cathode laser temporal profiles, Fig. \ref{PS_comp}, shows that the correlated energy spread and peak current values are achievable.  This comparison also hints at the benefits of temporal shaping of the cathode laser, perhaps suggesting a ramped or exponentially modified gaussian distribution, shifted in the head direction, similar to the results of \cite{penco2014experimental, PhysRevSTAB.9.120701}. 

\section{conclusion}
The analytical approach to backtracking outlined in this paper offers a fast, effective method to investigate the longitudinal phase space requirements at various points along the accelerator, relating the final and initial longitudinal phase spaces and current profiles with analytical expressions.  Comparison with forward tracking in Elegant shows that the method is robust and could be potentially applied to more exotic configurations.  The extension of this method to backtrack to the cathode is made complicated by the need to include transverse space charge effects and requires additional study.  However combining this with backtracking in the particle tracking code Opal \cite{adelmann2008opal}, offers a possible path forward.  

The method is limited in scope to piecewise continuous distributions, however in many cases the associated caustics are what we seek to avoid.  To avoid these caustics, a recommended starting point is to match the non-linear chirp acquired in the final drift/acceleration section, linearizing the phase space at the exit of the dispersive section. Given target final and initial phase spaces it may be possible to use this backtracking method to solve directly for an accelerator configuration, however we leave this for further study.  Furthermore, combining this backtracking solution with current machine learning techniques may be advantageous for swiftly finding a suitable initial phase space and accelerator configuration. 

\begin{acknowledgments}
The authors would like to thank Karl Bane, Yuri Nosochkov, Zhen Zhang, David Cesar, Joe Duris, Nicole Neveu and Arianna Formenti for useful discussions. This work was
supported by U.S. Department of Energy Contract No. DE-AC02-76SF00515 and award no. 2017-SLAC-100382.
\end{acknowledgments}

\begin{appendices}
\section{LCLS-II high current example}

\begin{figure}[h]
\centering
\includegraphics[scale=0.42]{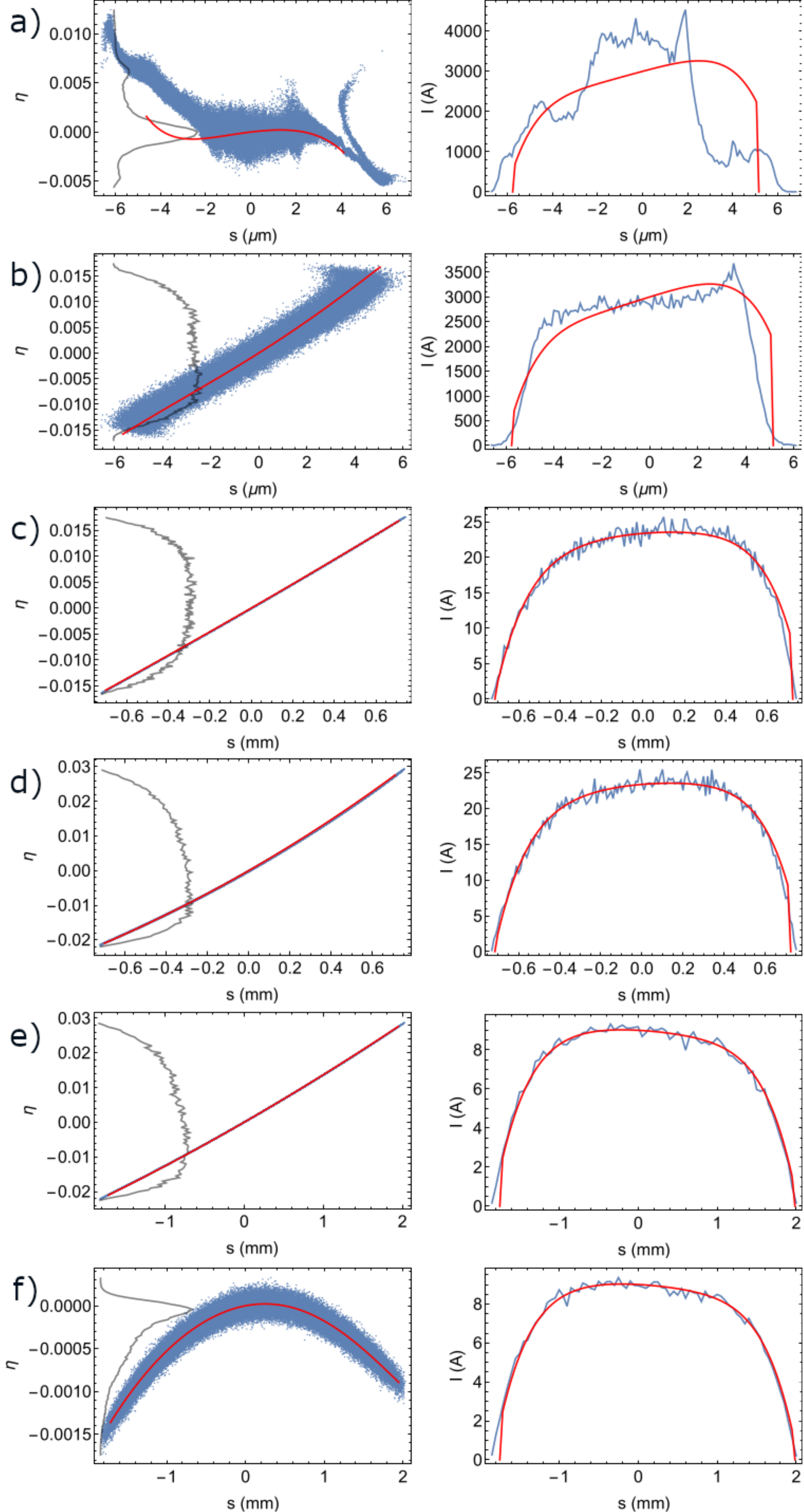}
\caption{Longitudinal phase space (left) and current profile (right) at the undulator entrance (a), BC2 exit (b), BC2 entrance (c), BC1 exit (d), BC1 entrance (e), and laser heater exit (f) with polynomials from backtracking (red) and particles from Elegant (blue).}
\label{BTRACKPS3}
\end{figure}

Here we include a second LCLS-II example case where we have tried to maximize the peak current at the undulator entrance.  In this case the evolution of the current profile in the bypass line cannot be ignored as can be seen in the comparison with forward tracking in Elegant, Fig. \ref{BTRACKPS3}.  The peak current at the BC2 exit is large enough to drive significant energy modulation from collective effects between dog leg bends producing additional dispersion and the chirp is large enough to experience compression in the compensating chicanes.  The accuracy of the backtracking method for this case could be improved by using Eq. 1-15 to solve for the evolution of the longitudinal phase space through each individual magnet and subsequent drift.  Regardless of this, forward tracking in Elegant of the longitudinal phase space and current profile at the laser heater exit found from backtracking gives a peak current of 4 kA with an energy spread of $\sigma_\eta =0.1 \%$ in the core of the beam.  Simulation parameters are given in Table II.  

\noindent
\begin{table}[ht]
\caption{LCLS-II example case 2 parameters}
\begin{tabular}{|c|c|c|c|}
\hline
\\[-1em]
$E_{0}$ $(MeV)$ & 92 & $I_{0}$ $(A)$ & 9\\
\hline
\\[-1em]
$E_{BC1}$ $(MeV)$ & 250 & $R_{56}^{(1)}$ $(mm)$ & -46.70\\
\\[-1em]
\hline
$E_{BC2}$ $(MeV)$ & 1500 & $I_{BC1}$ $(A)$ & 23.4\\
\\[-1em]
\hline
\\[-1em]
$E_{f}$ $(MeV)$ & 4000 & $R_{56}^{(2)}$ $(mm)$ & -42.97\\
\hline
\\[-1em]
$V_{L1B}$ $(MV)$ & 15.992 & $\phi_{L1B}$$\degree$ & -25.06 \\
\hline
\\[-1em]
$V_{L1H}$ $(MV)$ & 4.592 & $\phi_{L1H}$$\degree$ & -176.62 \\
\\[-1em]
\hline
$V_{L2B}$ $(MV)$ & 16.488 & $\phi_{L2B}$$\degree$ & -37.63 \\
\\[-1em]
\hline
\\[-1em]
$V_{L3B}$ $(MV)$ & 15.67 & $\phi_{L3B}$$\degree$ & 0 \\
\hline
\hline
\\[-1em]
$h_{f0}$ & 0 & $I_{f0}$ $(A)$ & 3000 \\
\hline
\\[-1em]
$h_{f1}$ $(m^{-1})$ & 281.94 & $I_{f1}$ & 50191.4 \\
\hline
\\[-1em]
$h_{f2}$ $(m^{-2})$ & -6.72$\times 10^7$ & $I_{f2}$ & -1.53$\times 10^9$ \\
\hline
\\[-1em]
$h_{f3}$ $(m^{-3})$ & -1.90$\times 10^{13}$  & $I_{f3}$ & -9.89$\times 10^{14}$ \\
\hline
\\[-1em]
$h_{f4}$ $(m^{-4})$ & 3.99$\times 10^{17}$ & $I_{f4}$ & -2.92$\times 10^{20}$ \\
\hline
\\[-1em]
$h_{f5}$ $(m^{-5})$ & -9.15$\times 10^{23}$  & $I_{f5}$ & 1.14$\times 10^{24}$\\
\hline
\\[-1em]
$h_{f6}$ $(m^{-6})$ & 3.49$\times 10^{28}$ & $I_{f6}$ & -9.31$\times 10^{30}$ \\
\hline
\\[-1em]
$h_{i0}$ & 0 & $I_{i0}$ $(A)$ & 9 \\
\hline
\\[-1em]
$h_{i1}$ $(m^{-1})$ & 0.176 & $I_{i1}$ &  -24.16 \\
\hline
\\[-1em]
$h_{i2}$ $(m^{-2})$ & -352.6 & $I_{i2}$ & -50166.5 \\
\hline
\\[-1em]
$h_{i3}$  $(m^{-3})$ & -4921.4 & $I_{i3}$ & -1.79$\times 10^{7}$ \\
\hline
\\[-1em]
$h_{i4}$  $(m^{-4})$ & -1.99$\times 10^{5}$ & $I_{i4}$  & -2.22$\times 10^{10}$ \\
\hline
\\[-1em]
$h_{i5}$  $(m^{-5})$ & 4.09$\times 10^{9}$ & $I_{i5}$ & 5.36$\times 10^{12}$ \\
\hline
\\[-1em]
$h_{i6}$  $(m^{-6})$ & 3.64$\times 10^{11}$ & $I_{i6}$ & -1.02$\times 10^{16}$ \\
\hline
\end{tabular}
\end{table}

\section{CSR chirp derivation}

The derivation of the CSR chirp follows \cite{saldin1997coherent,stupakov2002csr}, using the Lienard Wiechert potentials to find the electric field experienced by a particle as it traverses a bend magnet.  

First, for the calculations throughout, we flip the sign of the current distribution from Eq. 1 such that the head of the beam is on the right.

\begin{equation}
\begin{aligned}
&\bar{I}(s) = I_{0}(1-I_{1}s+I_{2}{s}^2-\\
&I_{3}{s}^3+\ldots+(-1)^N I_{N}{s}^N) \quad -S_2 < s < -S_1 \\
&\bar{I}(s) = 0 \quad s < -S_2\quad \& \quad s >-S_1
\end{aligned}
\end{equation}

For case A in \cite{saldin1997coherent}, where the observation point, $s$, is inside the magnet and the source point, $s'$, is outside of the magnet entrance, the CSR wake kernel is given by:
\begin{equation}
w_A[s-s']=-\frac{4}{\rho^2 \phi}\delta\bigg[\frac{s-s'}{\rho}-\frac{\phi^3}{6}\bigg]
\end{equation}
Here $\phi$ is the angle of the observation point relative to the magnet entrance.  The instantaneous change of energy is given by integrating over the source points:
\begin{equation}
\begin{aligned}
&\frac{1}{\rho}\frac{d\eta_A}{d\phi}=-\frac{e}{m c^3 4 \pi \epsilon_0 \gamma}\int ds' w_A[s-s']\bar{I}[s']\\
&=\frac{e}{m c^3 4 \pi \epsilon_0 \gamma}\frac{4}{\rho \phi}\bar{I}\bigg[s-\frac{\rho \phi^3}{6}\bigg]
\end{aligned}
\end{equation}
The total energy change is given by integrating over the observation point's path through the magnet up until the formation length of the radiation exceeds the tail of the beam.
\begin{equation}
\begin{aligned}
&\Delta \eta_A = \frac{e}{m c^3 4 \pi \epsilon_0 \gamma}\int_0^{\phi_f} d\phi \frac{4}{\phi}\bar{I}\bigg[s-\frac{\rho\phi^3}{6}\bigg]\\
&\phi_f = \bigg[ \frac{6}{\rho}(s+S_2)\bigg]^{1/3}
\end{aligned}
\end{equation}
For case B in \cite{saldin1997coherent}, where the observation point, $s$, is inside the magnet and the source point, $s'$ is inside the magnet, the CSR wake kernel is given by:
\begin{equation}
w_B[s-s']=-\frac{2}{(3\rho^2)^{1/3}}\frac{d}{ds'} \frac{1}{(s-s')^{1/3}}
\end{equation}
The instantaneous change of energy is given by integrating over the source points:
\begin{equation}
\begin{aligned}
&\frac{1}{\rho}\frac{d\eta_B}{d\phi}=-\frac{e}{m c^3 4 \pi \epsilon_0 \gamma}\int ds' w_B[s-s']\bar{I}[s']\\
&=-\frac{e}{m c^3 4 \pi \epsilon_0 \gamma}\bigg( \frac{4}{\rho \phi}\bar{I}\bigg[s-\frac{\rho \phi^3}{24}\bigg]+\\
&\frac{2}{(3\rho^2)^{1/3}}\int_{s-\frac{\rho \phi^3}{24}}^{s}ds' \frac{1}{(s-s')^{1/3}}\frac{d\bar{I}[s']}{ds'}\bigg)
\end{aligned}
\end{equation}
The total energy change is given by integrating over the observation point's path through the magnet up until the tail of the beam enters the magnet.  
\begin{equation}
\begin{aligned}
&\Delta \eta_B =-\frac{e}{m c^3 4 \pi \epsilon_0 \gamma}\int_0^{\phi_f} d\phi \bigg( \frac{4}{\phi}\bar{I}\bigg[s-\frac{\rho \phi^3}{24}\bigg]+\\
&\frac{2}{(3\rho^2)^{1/3}}\int_{s-\frac{\rho \phi^3}{24}}^{s}ds' \frac{1}{(s-s')^{1/3}}\frac{d\bar{I}[s']}{ds'}\bigg)\\
&\phi_f = \bigg[ \frac{24}{\rho}(s+S_2)\bigg]^{1/3}
\end{aligned}
\end{equation}
The sum of the integrals in Eq. 35 and 38 can be done explicitly, cancelling the logarithmic dependence at $\phi=0$.  Rewriting the energy modulation such that the head of the beam is again on the left, the polynomial chirp coefficients associated with this entrance transient are given by:
\begin{equation}
\begin{aligned}
&H_{A+B(n)} = -\frac{e(-1)^n}{m c^3 4 \pi \epsilon_0 \gamma}\times\\
&\bigg(\frac{4}{3}\log{(4)}(-1)^n\chi_n +\sum_{k=n}^{N}\sum_{i=0}^{n-1}\Xi_{n,k,i}(-1)^k\chi_k {S_2}^{k-n}+\\
&\sum_{k=n+1}^{N}\Xi_{n,k,n}(-1)^k\chi_k {S_2}^{k-n}\bigg)\\
\end{aligned}
\end{equation}
The coefficient, $\Xi$, is independent of the beam and magnet parameters and are  given by:
\begin{equation}
\Xi_{n,k,i}=\frac{8^{\frac{1}{3}} k(\frac{2}{3})^{(\overline{k-1-i})}(1-k)^{(\overline{k-1-i})}}{(k-n)!(n-i)!(\frac{5}{3})^{(\overline{k-1-i})}}
\end{equation}
Here $x^{\overline{n}}$ is the pochhammer function or rising factorial, $x^{\overline{n}}=x(x+1)(x+2)...(x+n-1)$.

Provided the beam satisfies the condition $\Phi > \bigg[ \frac{24}{\rho}(S_2-S_1)\bigg]^{1/3}$, the entire beam will experience a steady state CSR wake. For the case where this condition is not satisfied, some observation points in the beam will exit the magnet before the tail enters, case C in \cite{saldin1997coherent}.  In this regime, parts of the beam will experience a different CSR wake, complicating the analytical expression of the polynomial coefficients of the CSR chirp. Thus for simplicity, we assume the steady state condition is satisfied for the magnets in consideration. The CSR wake kernel for the steady state regime is given by the wake kernel for case B.
\begin{equation}
w_{SS}[s-s']=-\frac{2}{(3\rho^2)^{1/3}}\frac{d}{ds'} \frac{1}{(s-s')^{1/3}}
\end{equation}
The instantaneous change of energy is again given by integrating over the source points:
\begin{equation}
\begin{aligned}
&\frac{1}{\rho}\frac{d\eta_{SS}}{d\phi}=-\frac{e}{m c^3 4 \pi \epsilon_0 \gamma}\int ds' w_{SS}[s-s']\bar{I}[s']\\
&=-\frac{e}{m c^3 4 \pi \epsilon_0 \gamma}\bigg( \frac{4}{\rho \big[\frac{24}{\rho}(s+S_2)\big]^{1/3}}\bar{I}[-S_2]+\\
&\frac{2}{(3\rho^2)^{1/3}}\int_{-S_2}^{s}ds' \frac{1}{(s-s')^{1/3}}\frac{d\bar{I}[s']}{ds'}\bigg)
\end{aligned}
\end{equation}
The total energy change is given by the instantaneous energy change at the point the tail enters the magnet multiplied the distance travelled through the magnet until the the observation point exits.  
\begin{equation}
\begin{aligned}
&\Delta \eta_{SS} =-\frac{e}{m c^3 4 \pi \epsilon_0 \gamma}\rho (\phi_f - \phi_i) \bigg( \frac{4}{\rho \big[\frac{24}{\rho}(s+S_2)\big]^{1/3}}\bar{I}[-S_2]+\\
&\frac{2}{(3\rho^2)^{1/3}}\int_{-S_2}^{s}ds' \frac{1}{(s-s')^{1/3}}\frac{d\bar{I}[s']}{ds'}\bigg)\\
&\phi_i = \bigg[ \frac{24}{\rho}(s+S_2)\bigg]^{1/3} \quad \quad \phi_f =\Phi
\end{aligned}
\end{equation}
Again rewriting the energy modulation such that the head of the beam is on the left, the polynomial chirp coefficients associated with the steady state regime are given by:
\begin{equation}
\begin{aligned}
&H_{SS(n)} = -\frac{e(-1)^n}{m c^3 4 \pi \epsilon_0 \gamma}\times\\
&\bigg(\sum_{k=0}^{N}\Lambda_{n,k}^0 \Phi {\rho}^{\frac{1}{3}}(-1)^k\chi_k {S_2}^{k-n-\frac{1}{3}}+\\
&\sum_{m=0}^{n}\sum_{k=m+1}^{N}\sum_{i=0}^{m}\Lambda_{n,m,k,i}^1 \Phi {\rho}^{\frac{1}{3}}(-1)^k\chi_k {S_2}^{k-n-\frac{1}{3}}-\\
&\sum_{k=n+1}^{N}\sum_{i=0}^{m}\Lambda_{n,k,i}^2(-1)^k \chi_k {S_2}^{k-n}-\sum_{i=0}^{n-1}\Lambda_{n,n,i}^2(-1)^n\chi_n \bigg)
\end{aligned}
\end{equation}
The coefficients $\Lambda$ are independent of the beam and magnet parameters and are given by:
\begin{equation}
\begin{aligned}
&\Lambda_{n,k}^0=\frac{(-1)^k 4}{n! (24)^{\frac{1}{3}}}\bigg(-\frac{1}{3}\bigg)^{(\underline{n})}\\
&\Lambda_{n,m,k,i}^1=\frac{3^{\frac{2}{3}} k(\frac{2}{3})^{(\underline{n-m})}(\frac{2}{3})^{(\overline{k-1-i})}(1-k)^{(\overline{k-1-i})}}{(n-m)!(m-i)!(k-m-i)!(\frac{5}{3})^{(\overline{k-1-i})}}\\
&\Lambda_{n,k,i}^2=\frac{6 k(k-i)(\frac{2}{3})^{(\overline{k-1-i})}(1-k)^{(\overline{k-1-i})}}{(n-i)!(k-n)!(\frac{5}{3})^{(\overline{k-1-i})}}\\
\end{aligned}
\end{equation}
Here $x^{\overline{n}}$ is again the pochhammer function or rising factorial, and $y^{\underline{n}}$ is the falling factorial $y^{\underline{n}}=y(y-1)(y-2)...(y-n+1)$.

For case D in \cite{saldin1997coherent}, where the observation point, $s$, is outside the magnet exit and the source point, $s'$, is inside the magnet, the CSR wake kernel is given by:
\begin{equation}
w_{D}[s-s']= -\frac{4}{\rho}\frac{d}{ds'}\frac{1}{\psi[s'] +2x}
\end{equation}
Here $\psi$ is the angle of the source point from the magnet exit and $x$ is the distance of the observation point from the magnet exit.  The distance between source point and observation point is related by $\psi$ and $x$ by:
\begin{equation}
s-s' = \frac{\rho \psi^3}{24}\frac{\psi +4 x}{\psi +x}
\end{equation}
The instantaneous change of energy is given by integrating over the source points:
\begin{equation}
\begin{aligned}
&\frac{1}{\rho}\frac{d\eta_{D}}{d\phi}=-\frac{e}{m c^3 4 \pi \epsilon_0 \gamma}\int ds' w_D[s-s']\bar{I}[s']\\
&=-\frac{e}{m c^3 4 \pi \epsilon_0 \gamma}\frac{4}{\rho}\bigg( \frac{\bar{I}[-S_2]}{\psi[-S_2]+2x}+\\
&\int_{-S_2}^{s}ds' \frac{1}{\psi[s']+2x}\frac{d\bar{I}[s']}{ds'}\bigg)
\end{aligned}
\end{equation}
In order for us to solve the above integral we first define the angle of the tail of the beam, $\psi_0$, when the observation point is at distance $x$: 
\begin{equation}
s+S_2 = \frac{\rho {\psi_0}^3}{24}\frac{\psi_0 +4 x}{\psi_0 +x}
\end{equation}
With the above relation $x$ can be defined in terms of $\psi_0$.
\begin{equation}
x[\psi_0] = \frac{\rho {\psi_0}^4-24(s+S_2)\psi_0}{24(s+S_2)-4\rho{\psi_0}^3}
\end{equation}
The integral giving the instantaneous change of energy can be done changing variables and integrating over the angle of the source points:  
\begin{equation}
\begin{aligned}
&\frac{1}{\rho}\frac{d\eta_{D}}{d\phi}=-\frac{e}{m c^3 4 \pi \epsilon_0 \gamma}\frac{4}{\rho}\bigg( \frac{\bar{I}[-S_2]}{\psi_0+2x[\psi_0]}-\\
&\int_{0}^{\psi_0}d\psi \frac{1}{\psi+2x[\psi_0]}\frac{d}{d\psi}\bar{I}\bigg[s-\frac{\rho \psi^3}{24}\frac{\psi +4 x[\psi_0]}{\psi +x[\psi_0]}\bigg]\bigg)
\end{aligned}
\end{equation}
The total energy change is given by integrating over the angle of the beam tail:.
\begin{equation}
\begin{aligned}
&\Delta \eta_{D} =\frac{4e}{m c^3 4 \pi \epsilon_0 \gamma}\int_{\psi_0i}^{\psi_{0f}}d\psi_0 \bigg( \frac{\bar{I}[-S_2]}{\psi_0+2x[\psi_0]}-\\
&\int_{0}^{\psi_0}d\psi \frac{1}{\psi+2x[\psi_0]}\frac{d}{d\psi}\bar{I}\bigg[s-\frac{\rho \psi^3}{24}\frac{\psi +4 x[\psi_0]}{\psi +x[\psi_0]}\bigg]\bigg)\\
&\psi_{0i} = \bigg[ \frac{6}{\rho}(s+S_2)\bigg]^{1/3} \quad \quad \psi_{0f} = \bigg[ \frac{24}{\rho}(s+S_2)\bigg]^{1/3}
\end{aligned}
\end{equation}
In general the above integral can not be done analytically.  However, we can proceed by approximating the integral term in Eq. 51 associated with each polynomial coefficient of the current profile as being quadratic in $\psi_0$. Referring to these integral terms as, $F_{D(n)}$, we choose a quadratic approximation, $f_{D(n)}$:
\begin{equation}
\begin{aligned}
&F_{D(n)}[\psi_0]=-\int_{0}^{\psi_0}d\psi (-1)^n n \times\\
&\bigg(s-\frac{\rho \psi^3}{24}\frac{\psi +4 x[\psi_0]}{\psi +x[\psi_0]}\bigg)^{n-1}\frac{\rho \psi^2 (\psi+2x[\psi_0])}{8(\psi+x[\psi_0])^2}\\
&\sim f_{D(n)}[\psi_0] = A_{(n)}+B_{(n)}\bigg(\psi_0-\bigg[ \frac{24}{\rho}(s+S_2)\bigg]^{1/3}\bigg)+\\
&C_{(n)}\bigg(\psi_0-\bigg[ \frac{24}{\rho}(s+S_2)\bigg]^{1/3}\bigg)^2
\end{aligned}
\end{equation}
Here we choose $A_{(n)}$, $B_{(n)}$, and $C_{(n)}$ such that $F_{D(n)}=f_{D(n)}$ as x goes to 0, $f_{D(n)}=0$ as x goes to infinity, and $F_{D(n)}=f_{D(n)}$ at ${\psi_0}^3=\frac{1}{2}{\psi_{0f}}^3$:
\begin{equation}
\begin{aligned}
&A_{(n)}=F_{D(n)}\bigg[\bigg( \frac{24}{\rho}(s+S_2)\bigg)^{1/3}\bigg]\\
&f_{D(n)}\bigg[\bigg( \frac{6}{\rho}(s+S_2)\bigg)^{1/3}\bigg]=0\\
&f_{D(n)}\bigg[\bigg( \frac{12}{\rho}(s+S_2)\bigg)^{1/3}\bigg]=F_{D(n)}\bigg[\bigg( \frac{12}{\rho}(s+S_2)\bigg)^{1/3}\bigg]
\end{aligned}
\end{equation}
Solving the above system of equations and inserting $f_{D(n)}$ into Eq. 52, the integral is now trivial.  Again rewriting the energy modulation such that the head of the beam is on the left, the polynomial chirp coefficients associated with the exit transient are given by:
\begin{equation}
\begin{aligned}
&H_{D(n)} \sim -\frac{e(-1)^n}{m c^3 4 \pi \epsilon_0 \gamma}\times\\
&\bigg(\sum_{k=0}^{N-n}\sum_{m=0}^{n} (-1)^{k+n}\chi_{k+n}{S_2}^k \bigg[ \Upsilon_{n,m,k}^0 +\sum_{i=0}^{m+k-1}\Upsilon_{n,i,m,k}^1 -\\
&\sum_{i=0}^{m+k-1}\sum_{l=0}^{m+k-i-2}\Upsilon_{n,l,i,m,k}^2-\sum_{i=0}^{m+k-1}\sum_{l=m+k-i+1}^{3(m+k)-1}\Upsilon_{n,l,i,m,k}^2\bigg]\bigg)
\end{aligned}
\end{equation}
The coefficients, $\Upsilon$, are independent of the beam and magnet parameters and are given by:
\begin{equation}
\begin{aligned}
&\Upsilon_{n,m,k}^0=\frac{3(11-2^{\frac{1}{3}}10+2^{\frac{2}{3}})(-1)^{k+m}(k+m)!(k+n)!}{(2^{\frac{1}{3}}4-5)(3(k+m)-1)k!m!(k+m-1)!(n-m)!}\\
&\Upsilon_{n,i,m,k}^1=(-1)^{k+m-i}\times\\
&\frac{2^{\frac{1}{3}}12^{m+k-1}4[2i(4+\log{(27)})+(k+m)(10+\log{(27)})]}{2^{6(m+k)-1}3^i i! m! k!}\times\\
&\frac{(k+n)!(m+k)!(3 (k + m) - 1)!}{(k + m - 1 - i)! (n - m)! (k + m - i)! (i + 2 (k + m))!}\\
&\Upsilon_{n,l,i,m,k}^2 = \frac{(-1)^{l-1}}{(n-m)!}\bigg(\frac{3^{i+l-k-m}-1}{i+l-k-m} +\frac{3^{i+l-k-m+1}-1}{i+l-k-m+1}\bigg)\times\\
&\frac{2^{\frac{1}{3}}12^{m+k-1}4(k+n)!(k+m)!(3(k+m)-1)!}{2^{6(m+k)-1}3^{i-1}i!l!m!k!(k+m-1-i)!(3(k+m)-1-l)!}
\end{aligned}
\end{equation}
Combining the entrance transient, steady state and exit transients gives the polynomial coefficients of the CSR chirp. A list of bypass line magnets considered in the LCLS-II example case are given in Table III.
\noindent
\begin{table}[ht]
\caption{Bypass line magnet table}
\begin{tabular}{|c|c|c|}
\hline
\\[-1em]
$\Phi$ & $L_m$ & \# of magnets \\
\hline
\hline
\\[-1em]
0.01234 & 0.35 & 4 \\
\hline
\\[-1em]
0.02448 & 1.0 & 2 \\
\hline
\\[-1em]
0.01234 & 0.35 & 4 \\
\hline
\\[-1em]
0.00553  & 1.0 & 1 \\
\hline
\\[-1em]
0.00445 & 0.6 & 1 \\
\hline
\\[-1em]
0.0037 & 0.6 & 1 \\
\hline
\\[-1em]
0.0065 & 1.0 & 2 \\
\hline
\\[-1em]
0.00553  & 1.0 & 1 \\
\hline
\\[-1em]
0.00233  & 1.025 & 2 \\
\hline
\\[-1em]
0.00873  & 2.623 & 2 \\
\hline
\\[-1em]
0.0101  & 0.35 & 4 \\
\hline
\\[-1em]
0.00713  & 0.35 & 4 \\
\hline
\\[-1em]
0.00873  & 2.623 & 2 \\
\hline
\\[-1em]
0.00713  & 0.35 & 4 \\
\hline
\end{tabular}
\end{table}

\end{appendices}

\bibliographystyle{unsrt}
\bibliography{refs,An_analytical_backtracking_method_for_electron_beam_longitudinal_phase_space_shapingNotes}
\end{document}